\documentclass[aps,prl,showpacs,footinbib,superscriptaddress,twocolumn,longbibliography]{revtex4-2}

\usepackage{amsfonts}
\usepackage{amsmath}
\usepackage{multirow}
\usepackage{txfonts}
\usepackage{amssymb}
\usepackage{amsbsy}
\usepackage{graphicx}
\usepackage{epstopdf}
\usepackage{color}
\usepackage{braket} 
\usepackage{mathdots} 
\usepackage{booktabs}
\usepackage{indentfirst}
\usepackage{hyperref}
\usepackage{diagbox}
\usepackage{lipsum}
\hypersetup{hypertex=true, colorlinks=true, linkcolor=blue, anchorcolor=blue, citecolor=blue,urlcolor=blue}
\begin{document}
\title{Universal scaling of Green's functions in disordered non-Hermitian systems}	
\author{Yin-Quan Huang}
\altaffiliation{These authors contributed equally to this work.}
\affiliation{ Institute for Advanced Study, Tsinghua University, Beijing,  100084, China }

\author{Yu-Min Hu}
\altaffiliation{These authors contributed equally to this work.}
\affiliation{ Institute for Advanced Study, Tsinghua University, Beijing,  100084, China }

\author{Wen-Tan Xue}
\affiliation{ Institute for Advanced Study, Tsinghua University, Beijing,  100084, China }
\affiliation{ Department of Physics, National University of Singapore, Singapore 117542, Singapore }

\author{Zhong Wang}
\altaffiliation{ wangzhongemail@tsinghua.edu.cn }
\affiliation{ Institute for Advanced Study, Tsinghua University, Beijing,  100084, China }  
\date{\today}
\begin{abstract}	
The competition between non-Hermitian skin effect and Anderson localization leads to various intriguing phenomena concerning spectrums and wavefunctions. Here, we study the linear response of disordered non-Hermitian systems, which is precisely described by the Green's function. We show that the average maximum value of matrix elements of Green’s functions, which quantifies the maximal response against an external perturbation, exhibits different phases characterized by different scaling behaviors with respect to the system size. Whereas the exponential-growth phase is also seen in the translation-invariant systems, the algebraic-growth phase is unique to disordered non-Hermitian systems. We explain the numerical findings using the large deviation theory, which provides analytical insights into the algebraic scaling factors of non-Hermitian disordered Green’s functions. Furthermore, we show that these scaling behaviors can be observed in the steady states of disordered open quantum systems, offering a quantum-mechanical avenue for their experimental detection. Our work highlights an unexpected interplay between non-Hermitian skin effect and Anderson localization.
\end{abstract}
\maketitle
\emph{Introduction.--}Non-Hermitian physics has been attracting a great deal of attention \cite{Ashida2021,Bergholtz2021RMP}. In non-Hermitian systems, the intriguing localization mechanism, called the non-Hermitian skin effect (NHSE), can squeeze the bulk states to the edge under open boundary conditions (OBC) \cite{yao2018edge,yao2018chern,kunst2018biorthogonal,lee2018anatomy, Alvarez2018, Zhang2022ReivewOnNHSE, ding2022non, lin2023topological}. These exponentially boundary-localized bulk states dramatically revise the bulk-boundary correspondence for non-Hermitian systems \cite{xiao2020non,helbig2020generalized,Ghatak2019NHSE,Weidemann2020topological}, and underlie the conceptions of the generalized Brillouin zone and non-Bloch band theory \cite{yao2018edge,Yokomizo2019,Zhang2020correspondence}.

The interplay between non-Hermiticity and disorder has long been a subject of study. Early studies found that non-Hermiticity can delocalize the otherwise localized wavefunctions in disordered systems \cite{Hatano1996,Hatano1997vortex,Hatano1998non-Hermitian}. Recently, significant efforts have been dedicated to understanding the competing role of NHSE and Anderson localization in shaping wavefunctions and spectra in disordered non-Hermitian systems \cite{jiang2019interplay,longhi2019metal-insulator,liu2020generalized,liu2020non-Hermtian,liu2021exact_mobility_edges,longhi2021phase_transition,lin2022topological,liu2021localization,luo2021transfer,luo2022unifying,Kawabata2021nonunitary,longhi2021maryland,longhi2021spectral,li2022engineering,wang2024disorderinduced}. There is also a growing interest in exploring non-Hermitian topology with disorders \cite{longhi2019topological,tang2020topological_anderson,lin2022observation,cai2021localization,wang2021detecting,claes2021skin}.

Green's functions are indispensable in describing the linear response to external perturbations, serving as a standard probing technology in non-Hermitian experiments \cite{helbig2020generalized,wang2022non,Slim_2024}. The non-Hermitian Green's functions may exhibit unexpected phenomena due to the nontrivial interplay between NHSE and Anderson localization. However, unlike its counterpart in translation-invariant non-Hermitian systems \cite{Wanjura2019, xue2021simple, Borgnia2020nonhermitian, Zirnstein2021, Zirnstein2021exponential, hu2023green}, Green's function in disordered non-Hermitian systems is less explored. While progress has been achieved through perturbation theory in weak disorder \cite{Wanjura2021disorder} and qualitative studies of simple models \cite{Trefethen2001spectra,TrefethenEmbree+2020}, a general theory for Green's functions in disordered non-Hermitian systems remains elusive.

In this work, we unveil the universal scaling properties of non-Hermitian disordered Green's functions under OBC, defined as $G(E)=(E-H)^{-1}$ for a one-dimensional (1D) disordered non-Hermitian Hamiltonian $H$ and a complex frequency $E$. The imaginary part of $E$ is interpreted as an overall gain or loss added to $H$. To quantitatively characterize the scaling properties of disordered Green's functions, we primarily investigate the following quantity:
\begin{equation}
	G_m(E)\equiv\mathbb{E}[{\max_{i,j} |G_{ij}(E)|}].\label{eq:definition_Gm}
\end{equation}
This quantity defines the averaged maximum absolute value of matrix elements in $G(E)$, with $\mathbb{E}[\cdot]$ denoting averaging over disorder realizations. The quantity $G_m(E)$ describes the maximal response of disordered systems to an external perturbation. Hence, $G_m(E)>1$ implies potential amplification.

We show that $G_m(E)$ for a finite system of length $L$ exhibits three intriguing scaling phases as $L$ increases: the exponential-growth phase, the algebraic-growth phase, and the bounded phase. While the first and third phases have been identified in translation-invariant systems with NHSE \cite{xue2021simple}, the unexpected algebraic growth of $G_m(E)$ is unique to disordered non-Hermitian systems, arising as a result of the interplay between NHSE and Anderson localization. Note that an algebraic growth of the norm of Green's function has been numerically observed in the special bidiagonal case \cite{TrefethenEmbree+2020}; however, a theory is lacking even for this simplest case. We explain the scaling properties of $G_m(E)$ through large deviation theory, which provides a quantitative theory for the numerical findings.   
We also show that the scaling behaviors of Green's functions can manifest in the steady states of disordered open quantum systems. This setup provides a practical avenue for experimentally testing the predictions in our work.

\begin{figure}
	\centering
	\includegraphics[width=8.5cm]{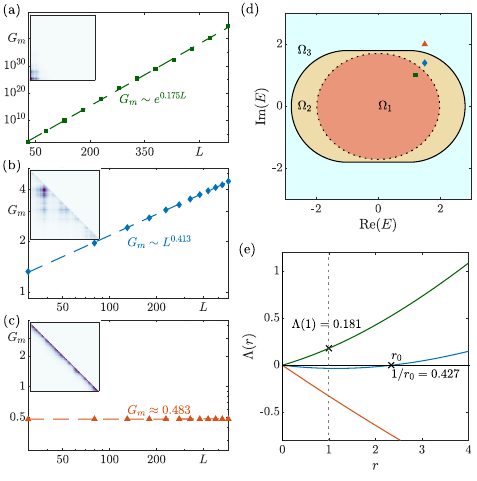}
	\caption{The scaling behaviors of $G_m(E)$ for $H_1$. We set $t=1.8$ and $V_i$ to be taken from a uniform distribution in the interval $[-1,1]$.  (a)--(c) show the scaling of $G_m(E)$ with $E=1.2+1i$ (a), $E=1.5+1.2i$ (b), and $E=1.5+2i$ (c). These data points are obtained by averaging over $10^4$ disorder realizations. The insets show a typical configuration of $G(E)$ with $L=80$, the darker color representing the larger $|G_{ij}(E)|$. (d) Phase diagram obtained from the large deviation theory with $\Omega_{1,2,3}$ corresponding to Table \ref{phasetable}. The dashed line is $\Lambda(1)=0$ and the solid line is $T_{\text{max}}=1$.  (e) Three typical $\Lambda(r)$ with $E$ taken from (a)--(c).  Theoretical results in (e) align well with numerical fittings in (a)--(b).}
 \label{fig:bi_phase}
\end{figure}

\emph{A simplest model.--} The three scaling behaviors of $G_m(E)$ can already be seen in a prototypical model with unidirectional hoppings and on-site disorders \cite{Trefethen2001spectra}. The Hamiltonian is
\begin{equation}
		H_{1}=\sum_{i=1}^{L-1}t c_{i+1}^\dagger c_i+\sum_{i=1}^LV_ic_i^\dagger c_i.\label{eq:bidiagonal}
\end{equation}
Here, $L$ is the system size and $t$ is the rightward unidirectional hopping. The potentials, $V_i$, are independent and identically distributed (i.i.d.) random variables. Specifically, for numerical simulations in Fig.\ref{fig:bi_phase}, we take a random potential uniformly distributed in the interval $[-1,1]$. As discussed below, our theory also applies to generic forms of disorders. The model in Eq.\eqref{eq:bidiagonal}, despite its simplicity, serves as a starting point for studying the scaling properties of Green's functions in general disordered non-Hermitian systems. 

Notably, the OBC Green's function $G(E)=(E-H_1)^{-1}$ with a specific realization of disorders can be obtained analytically. The matrix elements of $G(E)$ satisfy 
\begin{equation}
	(E-V_i)G_{ij}(E)-tG_{i-1,j}(E)=\delta_{ij}. \label{eq:bidiagonal_GF}
\end{equation}
 Since there are only rightward hoppings, $G_{ij}(E)=0$ for all $i<j$. Then for $i>j$, $G_{ij}(E)$  satisfies a recurrence relation $G_{ij}(E)=T_{i}(E)G_{i-1,j}(E)$ with transfer coefficient $T_{i}(E)=t/(E-V_i)$ and initial condition $G_{jj}(E)=1/(E-V_j)=T_j(E)/t$. Therefore, $G_{ij}(E)=(1/t)\prod_{k=j}^iT_k$ for $i\ge j$, where we use $T_i$ to represent the $E$-dependent transfer coefficients for brevity. For later convenience, we also define $T_{\text{max}}=\max_{V_i}|T_i|$ and $T_{\text{min}}=\min_{V_i}|T_i|$, which are independent of $i$. 

Intuitively, if $|T_i|\ge T_{\text{min}}>1$, $G_m(E)$ grows exponentially as $L$ increases. In contrast, if $|T_i|\le T_{\text{max}}<1$, $G_m(E)$ is bounded due to the exponential decay of $G_{ij}(E)$ with $i-j$. In the middle case of $T_{\text{max}}>1>T_{\text{min}}$, we find a lower bound as $G_m(E)\ge\mathbb{E}[{|G_{L1}(E)|}]=(1/t)\mathbb{E}[{\prod_{i=1}^L|T_i|}]=(1/t)\prod_{i=1}^L\mathbb{E}[{|T_i|}]=T_{\text{ave}}^{L}/t$ and an upper bound as $G_m(E)\le\mathbb{E}[{\sum_{ij}|G_{ij}(E)|}]=(1/t)\sum_{n=1}^L(L-n+1)T_{\text{ave}}^n=C_0+C_1 L+C_2 T^L_{\text{ave}}$, where $T_{\text{ave}}=\mathbb{E}[{|T_i|}]$ represents the averaged norm of the transfer coefficient and $C_{0,1,2}$ are $L$-independent constants. Therefore, if $\kappa\equiv\log T_{\text{ave}}>0$, $G_m(E)\sim\exp(\kappa L)$ indicates exponential amplification [Fig.\ref{fig:bi_phase}(a)].

Surprisingly, an intriguing algebraic growth of $G_m(E)$ is observed when $T_{\text{ave}}<1<T_{\text{max}}$ [Fig.\ref{fig:bi_phase}(b)]. In this case, it becomes possible for the product sequence $S_n=\prod_{i=1}^{n}T_i$, whose magnitude $|S_n|$ describes the growth or decay of $G_{j+n,j}(E)$, to initially grow with some $|T_i|>1$ and eventually decay for large $n$ due to self-averaging [see the inset of Fig. \ref{fig:bi_phase}(b)]. The self-averaging is captured by the Lyapunov exponent $\lambda\equiv\lim\limits_{n\to+\infty}\frac{1}{n}\log|S_n|=\mathbb{E}[{\log|T_i|}]$, which satisfies $\exp(\lambda)\le T_{\text{ave}}<1$ in this case. Intuitively, the possible length of a rare region containing a significant number of $|T_i|>1$, which supports the initial growth of $S_n$ for small $n$, is expected to increase with $L$. Therefore, we anticipate that $G_m(E)$ also grows with $L$, while $G_m(E)<C_1 L$ due to $T_{\text{ave}}<1$. This provides the physical origin of the sublinear algebraic scaling $G_m(E)\sim L^\alpha$ with $\alpha<1$ [Fig.\ref{fig:bi_phase}(b)].

\emph{Large deviation theory.--}With the insight into extreme fluctuations of $|S_n|$, we employ the large deviation theory in probability theory to analytically derive the scaling exponent $\alpha$ under the condition $T_{\text{ave}}<1<T_{\max}$. The large deviation theory serves as a framework for analyzing the asymptotic behavior of rare events or extreme fluctuations in random processes \cite{touchette2009large,touchette2012basic}. It finds extensive applications in statistical physics and Markovian dynamics \cite{ Giardina2006direct, Garrahan2007dynamical, Garrahan2010thermodynamics, Chetrite2013nonequilibrium, Znidaric2014exact, Carollo2019unraveling, chetrite2015nonequilibrium, jack2010large, hedges2009dynamic}.

Using the notation $\log|T|=\log|t/(E-V)|$ where $V$ is the same random variable as $V_i$, we define the \textit{cumulant generating function} of $\log|T|$ as $\Lambda(r)\equiv\log\left(\mathbb{E}[ {|T|^r}]\right)$.
Immediately, $T_{\text{ave}}=e^{\Lambda(1)}$ or $\kappa=\Lambda(1)$, indicating that a positive $\Lambda(1)$ encodes exponential growth $G_m(E)\sim e^{\Lambda(1)L}$ [Fig.\ref{fig:bi_phase}(a) and \ref{fig:bi_phase}(e)]. We then explore the case where $\Lambda(1)<0$. With the definition $\log|S_n|=\sum_{i=1}^n\log|T_i|$, we have shown that the self-averaging property results in $\lim\limits_{n\to+\infty}{|S_n|}\to 0$ since the Lyapunov exponent $\lambda=\mathbb{E}[{\log|T|}]<\Lambda(1)<0$. Given our focus on $G_m(E)$,  it becomes essential to determine the maximum possible value of $|S_n|$ for a finite system of length $L$. 

For a large but finite $n$, the large deviation theory indicates that the probability of finding $\frac{1}{n}\log|S_n|> s$ approximately follows the probability distribution $ P (\log|S_n|> ns)\approx e^{-nI(s)} $, where $I(s)\equiv\sup_{r>0}[sr-\Lambda(r)]$ is the Fenchel-Legendre transform of $\Lambda(r)$ \cite{touchette2012basic}. This result leads to

\begin{equation}\label{eq:large_deviation}
  \frac{1}{n}\log P \left(\log|S_n|> ns\right)\approx -I(s).
\end{equation}

The algebraic growth of $G_m(E)$ is explained by the extreme fluctuations of $|S_n|$ as follows. In a finite system of length $L$, a typical $|S_n|$ decays exponentially when $n$ is sufficiently large, which suggests that the possible maximum $|S_n|$ should be taken at a small length $n\ll L$. Consequently, the number of short $|S_n|$ sequences is proportional to $L$, which corresponds to the matrix elements $G_{i+n,i}(E)$ with $n/L\ll 1$ and  $1\le i\le L-n+1\approx L$. To let one of these $|S_n|$ sequences in a finite system approach a relatively large $ns$ with the probability of order unity, we require that, for each individual $|S_n|$, the large deviation probability $P(\log|S_n|>ns)$ is of the order of $1/L$. Inserting $ P $ into Eq. \eqref{eq:large_deviation} provides $(1/n)\log(1/L)\sim -I(s)$, which indeed indicates that the typical length of the largest $|S_n|$ found in a finite system is given by $n\sim (\log L)/I(s)\ll L$. The logarithmic scaling of $n$ for the largest possible $|S_n|$ is also verified by the numerical results presented in the supplemental material \cite{supplemental}. These results further support the above estimation. Therefore, with a fixed large-deviation parameter $s$, the magnitude of the possible largest $|S_n|$  in this finite system can be approximated as $\log|S_{n}|>ns\approx [s/I(s)]\log L$. To maximize $|S_{n}|$, a simple calculation shown in the  supplemental material \cite{supplemental} reveals that $\sup_s[{s}/{I(s)}]={1}/{r_0}$, where $r_0$ satisfies $\Lambda(r_0)=0$ and $1/r_0<1$. Given that $G_m(E)$ is proportional to the largest $|S_n|$ in a finite system, we obtain the following estimation:
\begin{equation}
	G_m(E)\sim e^{\sup_s\left[\frac{s}{I(s)}\right]\log L}\sim L^{\frac{1}{r_0}}.\label{eq:algeraic_scaling}
\end{equation}
The existence of $1/r_0<1$ is guaranteed by the condition $\Lambda (1)<1<T_{\max} $, which implies that extreme fluctuations with numerous $|T_i|>1$ can induce a large $|S_n|$ contributing to $G_m(E)$. In the end, we get the sublinear algebraic scaling of $G_m(E)$ when $T_{\text{ave}}<1<T_{\text{max}}$. This is a central result of our work. The scaling factor $\alpha=1/r_0$ fits well with the numerical findings [Figs.\ref{fig:bi_phase}(b) and \ref{fig:bi_phase}(e)].  
 
Furthermore, when $\Lambda(1)<0$, the absence of a solution $r_0>1$ for $\Lambda(r_0)=0$ implies $T_{\text{max}}<1$, which indicates the bounded phase of $G_m(E)$ [Fig.\ref{fig:bi_phase}(c)]. In conclusion, $\Lambda(r)$ encodes the three scaling phases of $G_m(E)$ for the model Eq.\eqref{eq:bidiagonal}. These results are summarized in Table \ref{phasetable}.

\begin{table}
	\centering
	\caption{Three scaling phases of $G_m(E)$.}\label{phasetable}
	\begin{tabular}{cccc}
		\toprule[1pt]
		Phases& Scaling behaviors   & Conditions on $\Lambda(r)$& $G_m(E)$ \\
		\hline
		$\Omega_{1}$&Exponential growth &$\Lambda(1)>0$  &$ e^{\Lambda(1)L}$ \\
		$\Omega_{2}$&Algebraic growth& $\Lambda(r_0)=0$ with $ r_0>1$ &$ L^{1/r_0}$ \\
		$\Omega_{3}$ & Bounded phase&$\Lambda(r)<0$ for $r\in(0,+\infty)$& Bounded \\
		\bottomrule[1pt]
	\end{tabular}
\end{table}

\begin{figure}
	\centering
	\includegraphics[width=8.5cm]{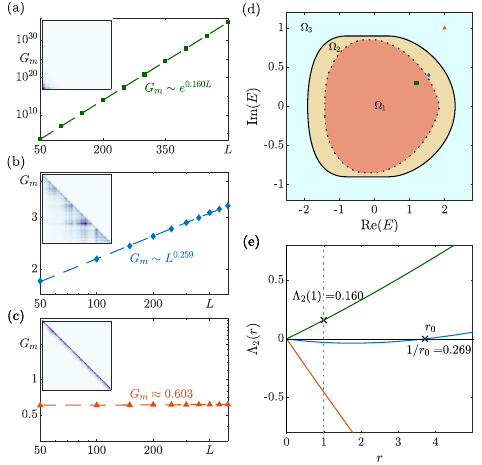}
    \caption{The scaling behaviors of $G_m(E)$ for $H_2$. We take $V_i$ from  a uniform distribution in the interval  $[-0.6,0.6]$ and set $(t_{-2},t_{-1},t_0,t_1,t_2)=(0.1,1.2,0.0,0.3,0.1)$.  (a)--(c) show the scaling of $G_m(E)$ with $E=1.2+0.3i$ (a), $E=1.55+0.4i$ (b), and $E=2+1i$ (c). These data points are obtained by averaging over $10^4$ disorder realizations. The insets show a typical configuration of $G(E)$ with $L=80$, with darker color meaning larger $|G_{ij}(E)|$. (d) The theoretical phase diagram with $\Omega_{1,2,3}$ related to Table \ref{phasetable}. The dashed line is $\Lambda_2(1)=0$ and the solid line is the marginal case without a nonzero root of $\Lambda_2(r)=0$.  (e) Three typical $\Lambda_2(r)$ with $E$ taken from (a)--(c).  Theoretical results in (e) align well with numerical fittings in (a)--(b).}
 \label{fig:general_model}
\end{figure}

\emph{General disordered systems.--}We now proceed to show that the three scaling behaviors of $G_m(E)$ universally survive in general disordered non-Hermitian systems, which can still be explained by the large deviation theory based on generalizing $\Lambda(r)$ [Fig.\ref{fig:general_model}]. Without loss of generality, we take the model with the next-nearest-neighbor hoppings as an example, whose Hamiltonian under OBC is 
\begin{equation}
	H_2=\sum_{i}\sum_{n=-2}^{2}t_nc_{i}^\dagger c_{i+n}+\sum_{i=1}^LV_ic_i^\dagger c_i.\label{eq:general_model}
\end{equation}
The translation-invariant part $H_0\equiv H_2|_{V_i=0}$ is generated by a non-Bloch Hamiltonian $h_0(\beta)=\sum_{n=-2}^{2}t_n\beta^n$ with $t_n$ being the hopping elements. The potentials $V_i$ are i.i.d. random variables following a specific distribution. Without loss of generality, we consider a uniform distribution in the interval  $[-0.6,0.6]$ in Fig.\ref{fig:general_model}. For simplicity, we assume that $H_0$ exhibits NHSE at the right boundary, indicating potential amplification in the right direction \cite{xue2021simple}. Consequently, to extract $G_m(E)$, we focus on $G_{ij}(E)$ with $i\ge j$, where the OBC Green's function is defined as $G(E)=(E-H_2)^{-1}$. The matrix elements of $G(E)$ satisfy

\begin{equation}
	(E-V_i)G_{ij}(E)-\sum_{n=-2}^2t_{n}G_{i+n,j}(E)=\delta_{ij}.\label{eq:general_transfer}
\end{equation}

We have seen that the transfer coefficient, defined below Eq.\eqref{eq:bidiagonal_GF}, plays a crucial role in determining the scaling factors of $G_m(E)$ for the  model Eq.\eqref{eq:bidiagonal}. For general disordered systems, the natural extension of the transfer coefficient is the transfer matrix \cite{kunst2018transfer,luo2021transfer,slevin2014critical}. Defining a vector $\psi_i^{(j)}(E)=(G_{i+1,j}(E),G_{i,j}(E),G_{i-1,j}(E),G_{i-2,j}(E))^T$ with $i>j$, the transfer matrix $T_i$ to the right direction is given by $\psi_{i+1}^{(j)}(E)=T_i(E)\psi_i^{(j)}(E)$:
\begin{equation}\label{eq:T_matrix}
	T_i(E)=\begin{pmatrix}
		-\frac{t_{1}}{t_2} & \frac{E-V_i-t_0}{t_2}& -\frac{t_{-1}}{t_2}  &-\frac{t_{-2}}{t_2}\\
		 1 & 0 & 0 & 0\\
		 0 & 1 & 0 & 0\\
	 	 0 & 0 & 1 & 0
	\end{pmatrix}.
\end{equation}
The matrices $T_i(E)$ are i.i.d. due to their dependence on i.i.d. random variables $V_i$. We omit $E$ below for brevity.

The product of $T_i$ generates a matrix sequence $S_n=\prod_{k=i}^{i+n} T_k$. Given that the diagonal elements of $G_{ij}(E)$ remain finite, $S_n$ encodes the scaling behavior of $G_{i+n,i}(E)$ as $n$ increases. By Oseledec's theorem, the exponential scaling behavior contained in $S_n$ is captured by the Lyapunov exponents (LEs) $\lambda_k$, defined as the eigenvalues of $\Omega=\lim\limits_{n\to\infty}\frac{1}{2n}\log(S_nS_n^\dagger)$. We order the LEs of Eq.\eqref{eq:T_matrix} as $\lambda_1\leq\cdots\leq\lambda_4$. 

The role of these LEs in disordered systems is parallel to that of the roots $\beta_k(E)$ (ordered by $|\beta_1(E)|\le|\beta_2(E)|\le|\beta_3(E)|\le|\beta_4(E)|$) of the characteristic equation $E=h_0(\beta)$ in the clean system with next-nearest-neighbor hoppings. In clean non-Hermitian systems, the non-Bloch band theory reveals that  $G_{ij}(E)=\braket{i|(E-H_0)^{-1}|j}=\oint_{\text{GBZ}}\frac{\mathrm{d}\beta}{2\pi i\beta}\frac{\beta^{i-j}}{E-h_0(\beta)}$, which is a contour integral along the generalized Brillouin zone (GBZ) \cite{xue2021simple,hu2023green}. In our clean model with next-nearest-neighbor hoppings, the GBZ encircles the roots $\beta_1(E)$ and $\beta_2(E)$ of $E=h_0(\beta)$. The asymptotic behavior of OBC Green's function is thus given by $|G_{i+n,i}(E)|\sim|\beta_2(E)|^n$. 

This motivates us to conjecture that Green's function in the disordered model (Eq.\eqref{eq:general_model}) scales as $|G_{i+n,i}(E)|\sim e^{n\lambda_2}$ for large $n$. This expectation is supported by extensive numerical simulations \cite{supplemental} and returns to the exact formula in clean systems when disorders are turned off. Immediately, $\lambda_2>0$ indicates the exponential amplification of $|G_{i+n,i}(E)|$.

However, if $\lambda_2<0$, we need to employ the large deviation theory to carefully investigate extreme fluctuations in a finite system, which requires the extension of the cumulant generating function $\Lambda(r)$ for the matrix sequence $S_n$. To facilitate this, we define a finite random sequence $\{\mathcal{S}_{k,1},\mathcal{S}_{k,2},\cdots,\mathcal{S}_{k,n}\}$ such that $\lim\limits_{n\to+\infty}\frac{1}{n}\log\mathcal{S}_{k,n}=\lambda_k$. This sequence naturally arises in the process of extracting LEs by QR decomposition \cite{slevin2014critical,luo2021transfer}. Specifically, considering a sequence of random matrices $\{T_1,T_2,\cdots,T_n,\cdots\}$ and an initial unitary matrix $Q_0$, we perform QR decomposition as $T_1Q_0=Q_1R_1$, where $Q_1$ is unitary and $R_1$ is upper triangular. This process is repeated recursively to obtain $T_nQ_{n-1}=Q_nR_n$ with a unitary $Q_n$ and an upper triangular $R_n$. Then the random sequence convergent to $e^{n\lambda_k}$ is defined as $\mathcal{S}_{k,n}\equiv|(\prod_{i=1}^nR_n)_{k,k}|=\prod_{i=1}^n|(R_n)_{k,k}|$. Given different realizations of $T_i$, $\mathcal{S}_{k,n}$ can be regarded as new random variables whose mean value provides $\lambda_k$ for a large $n$.

Based on the distribution of $\mathcal{S}_{k,n}$, the cumulant generating function is now defined as $\Lambda_k(r)=\lim\limits_{n\to\infty}\frac{1}{n}\log\mathbb{E}[{(\mathcal{S}_{k,n})^r}]$. Numerically, $\Lambda_k(r)$ is extracted from the vertical intercept $b$ of a linear fitting $\frac{1}{n}\log\mathbb{E}[{(\mathcal{S}_{k,n})^r}]=\frac{a}{n}+b$. For the model in Eq.\eqref{eq:general_model}, we are interested in the extreme fluctuations of $\mathcal{S}_{2,n}$, which correspond to the fluctuations of disordered Green's functions in the right direction. Therefore, the relevant cumulant generating function is $\Lambda(r)\equiv\Lambda_2(r)$. 

Similarly to the model in Eq.\eqref{eq:bidiagonal}, the presence of random sequences $\mathcal{S}_{2,n}$ with $\log\mathcal{S}_{2,n}>0$ indicates the potential growth of $G_m(E)$ as $L$ increases. Armed with $\Lambda(r)$ and following the same procedure as the discussions in the model Eq.\eqref{eq:bidiagonal}, we can readily apply the large deviation theory to the general model Eq.\eqref{eq:general_model} to investigate the scaling of $G_m(E)$. Depending on the complex $E$, we identify three scaling phases of $G_m(E)$ [Fig. \ref{fig:general_model}], identical to the results summarized in Table \ref{phasetable}. 

In conclusion, we have obtained a universal theory to describe the scaling behaviors of Green's functions in arbitrary 1D disordered non-Hermitian systems, including those with finite-range random hoppings or complex on-site random potentials. The key lies in extracting $\Lambda(r)$ from the corresponding transfer matrices $T_i$. The detailed theory is presented in the supplemental materials \cite{supplemental}.

\begin{figure}
	\centering
	\includegraphics[width=8.5cm]{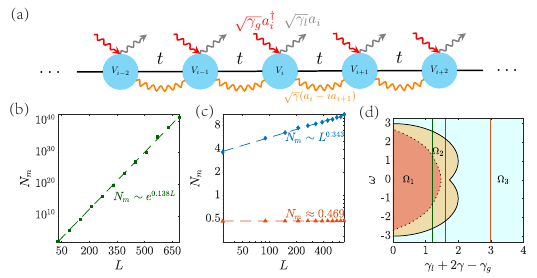}
	\caption{(a) The disordered open quantum system. We set $t=\gamma=\gamma_g=1$ and  $V_i$ to be taken from an equal binary distribution in $\{-1,1\}$.   (b) and (c) present the scaling of $N_m$ with $\gamma_l=0.2$ (green diamonds), $\gamma_l=0.6$ (blue squares), and $\gamma_l=2$ (orange triangles). These results are obtained from averaging over $10^4$ disorder realizations. (d) The phase diagram of $G_m(i\omega)$. We fix $\gamma_l+2\gamma-\gamma_g>0$ to make sure that the open quantum system has stable steady states. Three colored lines correspond to the parameters used in (b) and (c). }\label{fig:opensystem}
\end{figure}

\emph{Disordered open quantum systems.--}The Green's functions arise in various physical contexts, making our prediction testable in many experiments. Here, we propose a bosonic open quantum system where the scaling properties of Green's functions are reflected on the non-equilibrium steady states under open boundary conditions.   

We consider a disordered bosonic Lindblad master equation $
\frac{\mathrm{d}}{\mathrm{d} t} \rho(t)=-i\left[H_c,\rho(t)\right]+\sum_{\mu}(L_{\mu} \rho(t) L_{\mu}^{\dagger}-\frac{1}{2}\{L_{\mu}^{\dagger} L_{\mu}, \rho(t)\})$, shown in Fig. \ref{fig:opensystem}(a). Here, $\rho(t)$ is the density matrix of a system containing $L$ bosonic modes. The coherent Hamiltonian $H_c$ under OBC is given by $H_c=\sum_{ij}b_i^\dagger h_{ij}b_j=\sum_{i=1}^{L-1}t(b_i^\dagger b_{i+1}+b_{i+1}^\dagger b_{i})+\sum_{i=1}^LV_ib_i^\dagger b_i$, where $b_i$ are bosonic annihilation operators and $V_i$ are i.i.d. random potentials. The gain and loss jump operators, denoted as $L_\mu^{(g)}=\sum_i D_{\mu,i}^{(g)}b_i^\dagger$ and $L_\mu^{(l)}=\sum_i D_{\mu,i}^{(l)}b_i$ respectively, are given by the following three groups: (i) $L_i^{(g)}=\sqrt{\gamma_g}b_i^\dagger$; (ii) $L_i^{(l)}=\sqrt{\gamma_l}b_i$; (iii) $L_{i,i+1}^{(l)}=\sqrt{\gamma}(b_i-ib_{i+1})$. The last group, together with the translation-invariant part of $H_c$, generates NHSE in this open quantum system \cite{song2019chiral}.

The steady-state density distribution is proven to be $N_{\text{ss},i}=\operatorname{Tr}[\rho_{\text{ss}}b_i^\dagger b_i]=\gamma_{g} \sum_{j=1}^L\int_{-\infty}^{+\infty} \frac{\mathrm{d} \omega}{2\pi}|\langle i|(i \omega-X)^{-1}| j\rangle|^{2}$, where $\rho_{\text{ss}}$ is the steady-state density matrix and the damping matrix $X$ is $X=ih^T+\frac{1}{2}(M_g-(M_l)^T)$ with $M_g=D^{(g)\dagger} D^{(g)}=\gamma_g\mathbb{I}$ and $M_l=D^{(l)\dagger} D^{(l)}$ \cite{McDonald2022nonequilibrium,hu2023manybody}. This formula demonstrates that $N_{\text{ss},i}$ encodes the information of Green's functions $G(i\omega)= (i\omega-X)^{-1}$ and, consequently, reflects the scaling of $G_m(i\omega)$ as defined in Eq.\eqref{eq:definition_Gm} for $G(i\omega)$.

Without disorders, the NHSE in $X$ indicates that the maximum steady-state particle number $N_m\equiv\max_i (N_{\text{ss},i})$ grows exponentially with $L$ in the strong pumping region $\gamma_l<\gamma_g<2\gamma+\gamma_l$, while it remains bounded in the strong loss region $\gamma_g<\gamma_l$ \cite{McDonald2022nonequilibrium}. In the presence of disorders, $N_m$ shows three scaling phases: the exponential-growth phase, the algebraic-growth phase, and the bounded phase [Fig. \ref{fig:opensystem}], similar to the scaling behaviors in Table \ref{phasetable}. Although extracting the scaling factors of $N_m$ poses a challenge, an endeavor left for future investigation, we stress that the scaling of $N_m$ in open quantum systems shares the same origin as that of $G_m(i\omega)$. Fig. \ref{fig:opensystem}(d) demonstrates that the scaling behaviors of $N_m$ and $G_m(i\omega)$ are related through the parameters that control the integral path of $i\omega$ to cross different scaling regions in the phase diagram of $G_m(i\omega)$.

\emph{Discussions.--}In conclusion, utilizing the large deviation theory, we unraveled the universal scaling behaviors of Green's function in disordered non-Hermitian systems [Table \ref{phasetable}]. These unexpected scaling properties can be observed in a wide class of open quantum systems. Our theory offers a fresh perspective on the interplay between non-Hermiticity and disorders. Given the growing interest in non-Hermitian many-body localization \cite{Hamazaki2019nonhermitian, zhai2020many-body, Tang2021Localization, Heuen2021extracting, Suthar2022non-Hermitian, Wang2023nonhemitian, obrien2023probing,roccati2024diagnosing}, it is interesting to extend the large deviation theory into these non-Hermitian interacting systems. Another intriguing direction is to uncover the universal scaling properties of disordered Green's functions in high-dimensional non-Hermitian systems \cite{zhang2022universal, Jiang2023Dimensional, wang2022amoeba}.

\emph{Acknowledgment.--} This work is supported by the National Natural Science Foundation of China under
Grant No. 12125405, National Key R\&D Program of China (No. 2023YFA1406702), and
the Innovation Program for Quantum Science and Technology (No. 2021ZD0302502).
\bibliography{GF}

\begin{thebibliography}{79}%
\makeatletter
\providecommand \@ifxundefined [1]{%
 \@ifx{#1\undefined}
}%
\providecommand \@ifnum [1]{%
 \ifnum #1\expandafter \@firstoftwo
 \else \expandafter \@secondoftwo
 \fi
}%
\providecommand \@ifx [1]{%
 \ifx #1\expandafter \@firstoftwo
 \else \expandafter \@secondoftwo
 \fi
}%
\providecommand \natexlab [1]{#1}%
\providecommand \enquote  [1]{``#1''}%
\providecommand \bibnamefont  [1]{#1}%
\providecommand \bibfnamefont [1]{#1}%
\providecommand \citenamefont [1]{#1}%
\providecommand \href@noop [0]{\@secondoftwo}%
\providecommand \href [0]{\begingroup \@sanitize@url \@href}%
\providecommand \@href[1]{\@@startlink{#1}\@@href}%
\providecommand \@@href[1]{\endgroup#1\@@endlink}%
\providecommand \@sanitize@url [0]{\catcode `\\12\catcode `\$12\catcode
  `\&12\catcode `\#12\catcode `\^12\catcode `\_12\catcode `\%12\relax}%
\providecommand \@@startlink[1]{}%
\providecommand \@@endlink[0]{}%
\providecommand \url  [0]{\begingroup\@sanitize@url \@url }%
\providecommand \@url [1]{\endgroup\@href {#1}{\urlprefix }}%
\providecommand \urlprefix  [0]{URL }%
\providecommand \Eprint [0]{\href }%
\providecommand \doibase [0]{https://doi.org/}%
\providecommand \selectlanguage [0]{\@gobble}%
\providecommand \bibinfo  [0]{\@secondoftwo}%
\providecommand \bibfield  [0]{\@secondoftwo}%
\providecommand \translation [1]{[#1]}%
\providecommand \BibitemOpen [0]{}%
\providecommand \bibitemStop [0]{}%
\providecommand \bibitemNoStop [0]{.\EOS\space}%
\providecommand \EOS [0]{\spacefactor3000\relax}%
\providecommand \BibitemShut  [1]{\csname bibitem#1\endcsname}%
\let\auto@bib@innerbib\@empty
\bibitem [{\citenamefont {Ashida}\ \emph {et~al.}(2020)\citenamefont {Ashida},
  \citenamefont {Gong},\ and\ \citenamefont {Ueda}}]{Ashida2021}%
  \BibitemOpen
  \bibfield  {author} {\bibinfo {author} {\bibfnamefont {Y.}~\bibnamefont
  {Ashida}}, \bibinfo {author} {\bibfnamefont {Z.}~\bibnamefont {Gong}},\ and\
  \bibinfo {author} {\bibfnamefont {M.}~\bibnamefont {Ueda}},\ }\bibfield
  {title} {\bibinfo {title} {Non-hermitian physics},\ }\href
  {https://doi.org/10.1080/00018732.2021.1876991} {\bibfield  {journal}
  {\bibinfo  {journal} {Advances in Physics}\ }\textbf {\bibinfo {volume}
  {69}},\ \bibinfo {pages} {249} (\bibinfo {year} {2020})}\BibitemShut
  {NoStop}%
\bibitem [{\citenamefont {Bergholtz}\ \emph {et~al.}(2021)\citenamefont
  {Bergholtz}, \citenamefont {Budich},\ and\ \citenamefont
  {Kunst}}]{Bergholtz2021RMP}%
  \BibitemOpen
  \bibfield  {author} {\bibinfo {author} {\bibfnamefont {E.~J.}\ \bibnamefont
  {Bergholtz}}, \bibinfo {author} {\bibfnamefont {J.~C.}\ \bibnamefont
  {Budich}},\ and\ \bibinfo {author} {\bibfnamefont {F.~K.}\ \bibnamefont
  {Kunst}},\ }\bibfield  {title} {\bibinfo {title} {Exceptional topology of
  non-hermitian systems},\ }\href
  {https://doi.org/10.1103/RevModPhys.93.015005} {\bibfield  {journal}
  {\bibinfo  {journal} {Rev. Mod. Phys.}\ }\textbf {\bibinfo {volume} {93}},\
  \bibinfo {pages} {015005} (\bibinfo {year} {2021})}\BibitemShut {NoStop}%
\bibitem [{\citenamefont {Yao}\ and\ \citenamefont {Wang}(2018)}]{yao2018edge}%
  \BibitemOpen
  \bibfield  {author} {\bibinfo {author} {\bibfnamefont {S.}~\bibnamefont
  {Yao}}\ and\ \bibinfo {author} {\bibfnamefont {Z.}~\bibnamefont {Wang}},\
  }\bibfield  {title} {\bibinfo {title} {Edge states and topological invariants
  of non-hermitian systems},\ }\href
  {https://doi.org/10.1103/PhysRevLett.121.086803} {\bibfield  {journal}
  {\bibinfo  {journal} {Phys. Rev. Lett.}\ }\textbf {\bibinfo {volume} {121}},\
  \bibinfo {pages} {086803} (\bibinfo {year} {2018})}\BibitemShut {NoStop}%
\bibitem [{\citenamefont {Yao}\ \emph {et~al.}(2018)\citenamefont {Yao},
  \citenamefont {Song},\ and\ \citenamefont {Wang}}]{yao2018chern}%
  \BibitemOpen
  \bibfield  {author} {\bibinfo {author} {\bibfnamefont {S.}~\bibnamefont
  {Yao}}, \bibinfo {author} {\bibfnamefont {F.}~\bibnamefont {Song}},\ and\
  \bibinfo {author} {\bibfnamefont {Z.}~\bibnamefont {Wang}},\ }\bibfield
  {title} {\bibinfo {title} {Non-hermitian chern bands},\ }\href
  {https://doi.org/10.1103/PhysRevLett.121.136802} {\bibfield  {journal}
  {\bibinfo  {journal} {Phys. Rev. Lett.}\ }\textbf {\bibinfo {volume} {121}},\
  \bibinfo {pages} {136802} (\bibinfo {year} {2018})}\BibitemShut {NoStop}%
\bibitem [{\citenamefont {Kunst}\ \emph {et~al.}(2018)\citenamefont {Kunst},
  \citenamefont {Edvardsson}, \citenamefont {Budich},\ and\ \citenamefont
  {Bergholtz}}]{kunst2018biorthogonal}%
  \BibitemOpen
  \bibfield  {author} {\bibinfo {author} {\bibfnamefont {F.~K.}\ \bibnamefont
  {Kunst}}, \bibinfo {author} {\bibfnamefont {E.}~\bibnamefont {Edvardsson}},
  \bibinfo {author} {\bibfnamefont {J.~C.}\ \bibnamefont {Budich}},\ and\
  \bibinfo {author} {\bibfnamefont {E.~J.}\ \bibnamefont {Bergholtz}},\
  }\bibfield  {title} {\bibinfo {title} {Biorthogonal bulk-boundary
  correspondence in non-hermitian systems},\ }\href
  {https://doi.org/10.1103/PhysRevLett.121.026808} {\bibfield  {journal}
  {\bibinfo  {journal} {Phys. Rev. Lett.}\ }\textbf {\bibinfo {volume} {121}},\
  \bibinfo {pages} {026808} (\bibinfo {year} {2018})}\BibitemShut {NoStop}%
\bibitem [{\citenamefont {Lee}\ and\ \citenamefont
  {Thomale}(2019)}]{lee2018anatomy}%
  \BibitemOpen
  \bibfield  {author} {\bibinfo {author} {\bibfnamefont {C.~H.}\ \bibnamefont
  {Lee}}\ and\ \bibinfo {author} {\bibfnamefont {R.}~\bibnamefont {Thomale}},\
  }\bibfield  {title} {\bibinfo {title} {Anatomy of skin modes and topology in
  non-hermitian systems},\ }\href {https://doi.org/10.1103/PhysRevB.99.201103}
  {\bibfield  {journal} {\bibinfo  {journal} {Phys. Rev. B}\ }\textbf {\bibinfo
  {volume} {99}},\ \bibinfo {pages} {201103} (\bibinfo {year}
  {2019})}\BibitemShut {NoStop}%
\bibitem [{\citenamefont {Martinez~Alvarez}\ \emph {et~al.}(2018)\citenamefont
  {Martinez~Alvarez}, \citenamefont {Barrios~Vargas}, \citenamefont
  {Berdakin},\ and\ \citenamefont {Foa~Torres}}]{Alvarez2018}%
  \BibitemOpen
  \bibfield  {author} {\bibinfo {author} {\bibfnamefont {V.~M.}\ \bibnamefont
  {Martinez~Alvarez}}, \bibinfo {author} {\bibfnamefont {J.~E.}\ \bibnamefont
  {Barrios~Vargas}}, \bibinfo {author} {\bibfnamefont {M.}~\bibnamefont
  {Berdakin}},\ and\ \bibinfo {author} {\bibfnamefont {L.~E.~F.}\ \bibnamefont
  {Foa~Torres}},\ }\bibfield  {title} {\bibinfo {title} {Topological states of
  non-hermitian systems},\ }\href@noop {} {\bibfield  {journal} {\bibinfo
  {journal} {Eur. Phys. J. Spec. Top.}\ }\textbf {\bibinfo {volume} {227}},\
  \bibinfo {pages} {1295} (\bibinfo {year} {2018})}\BibitemShut {NoStop}%
\bibitem [{\citenamefont {Zhang}\ \emph
  {et~al.}(2022{\natexlab{a}})\citenamefont {Zhang}, \citenamefont {Zhang},
  \citenamefont {Lu},\ and\ \citenamefont {Chen}}]{Zhang2022ReivewOnNHSE}%
  \BibitemOpen
  \bibfield  {author} {\bibinfo {author} {\bibfnamefont {X.}~\bibnamefont
  {Zhang}}, \bibinfo {author} {\bibfnamefont {T.}~\bibnamefont {Zhang}},
  \bibinfo {author} {\bibfnamefont {M.-H.}\ \bibnamefont {Lu}},\ and\ \bibinfo
  {author} {\bibfnamefont {Y.-F.}\ \bibnamefont {Chen}},\ }\bibfield  {title}
  {\bibinfo {title} {A review on non-hermitian skin effect},\ }\href
  {https://doi.org/10.1080/23746149.2022.2109431} {\bibfield  {journal}
  {\bibinfo  {journal} {Advances in Physics: X}\ }\textbf {\bibinfo {volume}
  {7}},\ \bibinfo {pages} {2109431} (\bibinfo {year}
  {2022}{\natexlab{a}})}\BibitemShut {NoStop}%
\bibitem [{\citenamefont {Ding}\ \emph {et~al.}(2022)\citenamefont {Ding},
  \citenamefont {Fang},\ and\ \citenamefont {Ma}}]{ding2022non}%
  \BibitemOpen
  \bibfield  {author} {\bibinfo {author} {\bibfnamefont {K.}~\bibnamefont
  {Ding}}, \bibinfo {author} {\bibfnamefont {C.}~\bibnamefont {Fang}},\ and\
  \bibinfo {author} {\bibfnamefont {G.}~\bibnamefont {Ma}},\ }\bibfield
  {title} {\bibinfo {title} {Non-hermitian topology and exceptional-point
  geometries},\ }\href {https://doi.org/10.1038/s42254-022-00516-5} {\bibfield
  {journal} {\bibinfo  {journal} {Nature Reviews Physics}\ ,\ \bibinfo {pages}
  {1}} (\bibinfo {year} {2022})}\BibitemShut {NoStop}%
\bibitem [{\citenamefont {Lin}\ \emph {et~al.}(2023)\citenamefont {Lin},
  \citenamefont {Tai}, \citenamefont {Li},\ and\ \citenamefont
  {Lee}}]{lin2023topological}%
  \BibitemOpen
  \bibfield  {author} {\bibinfo {author} {\bibfnamefont {R.}~\bibnamefont
  {Lin}}, \bibinfo {author} {\bibfnamefont {T.}~\bibnamefont {Tai}}, \bibinfo
  {author} {\bibfnamefont {L.}~\bibnamefont {Li}},\ and\ \bibinfo {author}
  {\bibfnamefont {C.~H.}\ \bibnamefont {Lee}},\ }\bibfield  {title} {\bibinfo
  {title} {Topological non-hermitian skin effect},\ }\bibfield  {journal}
  {\bibinfo  {journal} {Frontiers of Physics}\ }\textbf {\bibinfo {volume}
  {18}},\ \href {https://doi.org/10.1007/s11467-023-1309-z}
  {10.1007/s11467-023-1309-z} (\bibinfo {year} {2023})\BibitemShut {NoStop}%
\bibitem [{\citenamefont {Xiao}\ \emph {et~al.}(2020)\citenamefont {Xiao},
  \citenamefont {Deng}, \citenamefont {Wang}, \citenamefont {Zhu},
  \citenamefont {Wang}, \citenamefont {Yi},\ and\ \citenamefont
  {Xue}}]{xiao2020non}%
  \BibitemOpen
  \bibfield  {author} {\bibinfo {author} {\bibfnamefont {L.}~\bibnamefont
  {Xiao}}, \bibinfo {author} {\bibfnamefont {T.}~\bibnamefont {Deng}}, \bibinfo
  {author} {\bibfnamefont {K.}~\bibnamefont {Wang}}, \bibinfo {author}
  {\bibfnamefont {G.}~\bibnamefont {Zhu}}, \bibinfo {author} {\bibfnamefont
  {Z.}~\bibnamefont {Wang}}, \bibinfo {author} {\bibfnamefont {W.}~\bibnamefont
  {Yi}},\ and\ \bibinfo {author} {\bibfnamefont {P.}~\bibnamefont {Xue}},\
  }\bibfield  {title} {\bibinfo {title} {Non-hermitian bulk–boundary
  correspondence in quantum dynamics},\ }\href
  {https://doi.org/10.1038/s41567-020-0836-6} {\bibfield  {journal} {\bibinfo
  {journal} {Nature Physics}\ }\textbf {\bibinfo {volume} {16}},\ \bibinfo
  {pages} {761–766} (\bibinfo {year} {2020})}\BibitemShut {NoStop}%
\bibitem [{\citenamefont {Helbig}\ \emph {et~al.}(2020)\citenamefont {Helbig},
  \citenamefont {Hofmann}, \citenamefont {Imhof}, \citenamefont {Abdelghany},
  \citenamefont {Kiessling}, \citenamefont {Molenkamp}, \citenamefont {Lee},
  \citenamefont {Szameit}, \citenamefont {Greiter},\ and\ \citenamefont
  {Thomale}}]{helbig2020generalized}%
  \BibitemOpen
  \bibfield  {author} {\bibinfo {author} {\bibfnamefont {T.}~\bibnamefont
  {Helbig}}, \bibinfo {author} {\bibfnamefont {T.}~\bibnamefont {Hofmann}},
  \bibinfo {author} {\bibfnamefont {S.}~\bibnamefont {Imhof}}, \bibinfo
  {author} {\bibfnamefont {M.}~\bibnamefont {Abdelghany}}, \bibinfo {author}
  {\bibfnamefont {T.}~\bibnamefont {Kiessling}}, \bibinfo {author}
  {\bibfnamefont {L.}~\bibnamefont {Molenkamp}}, \bibinfo {author}
  {\bibfnamefont {C.}~\bibnamefont {Lee}}, \bibinfo {author} {\bibfnamefont
  {A.}~\bibnamefont {Szameit}}, \bibinfo {author} {\bibfnamefont
  {M.}~\bibnamefont {Greiter}},\ and\ \bibinfo {author} {\bibfnamefont
  {R.}~\bibnamefont {Thomale}},\ }\bibfield  {title} {\bibinfo {title}
  {Generalized bulk--boundary correspondence in non-hermitian topolectrical
  circuits},\ }\href {https://doi.org/10.1038/s41567-020-0922-9} {\bibfield
  {journal} {\bibinfo  {journal} {Nature Physics}\ }\textbf {\bibinfo {volume}
  {16}},\ \bibinfo {pages} {747} (\bibinfo {year} {2020})}\BibitemShut
  {NoStop}%
\bibitem [{\citenamefont {Ghatak}\ \emph {et~al.}(2020)\citenamefont {Ghatak},
  \citenamefont {Brandenbourger}, \citenamefont {van Wezel},\ and\
  \citenamefont {Coulais}}]{Ghatak2019NHSE}%
  \BibitemOpen
  \bibfield  {author} {\bibinfo {author} {\bibfnamefont {A.}~\bibnamefont
  {Ghatak}}, \bibinfo {author} {\bibfnamefont {M.}~\bibnamefont
  {Brandenbourger}}, \bibinfo {author} {\bibfnamefont {J.}~\bibnamefont {van
  Wezel}},\ and\ \bibinfo {author} {\bibfnamefont {C.}~\bibnamefont
  {Coulais}},\ }\bibfield  {title} {\bibinfo {title} {Observation of
  non-hermitian topology and its bulk--edge correspondence in an active
  mechanical metamaterial},\ }\href {https://doi.org/10.1073/pnas.2010580117}
  {\bibfield  {journal} {\bibinfo  {journal} {Proceedings of the National
  Academy of Sciences}\ }\textbf {\bibinfo {volume} {117}},\ \bibinfo {pages}
  {29561} (\bibinfo {year} {2020})}\BibitemShut {NoStop}%
\bibitem [{\citenamefont {Weidemann}\ \emph {et~al.}(2020)\citenamefont
  {Weidemann}, \citenamefont {Kremer}, \citenamefont {Helbig}, \citenamefont
  {Hofmann}, \citenamefont {Stegmaier}, \citenamefont {Greiter}, \citenamefont
  {Thomale},\ and\ \citenamefont {Szameit}}]{Weidemann2020topological}%
  \BibitemOpen
  \bibfield  {author} {\bibinfo {author} {\bibfnamefont {S.}~\bibnamefont
  {Weidemann}}, \bibinfo {author} {\bibfnamefont {M.}~\bibnamefont {Kremer}},
  \bibinfo {author} {\bibfnamefont {T.}~\bibnamefont {Helbig}}, \bibinfo
  {author} {\bibfnamefont {T.}~\bibnamefont {Hofmann}}, \bibinfo {author}
  {\bibfnamefont {A.}~\bibnamefont {Stegmaier}}, \bibinfo {author}
  {\bibfnamefont {M.}~\bibnamefont {Greiter}}, \bibinfo {author} {\bibfnamefont
  {R.}~\bibnamefont {Thomale}},\ and\ \bibinfo {author} {\bibfnamefont
  {A.}~\bibnamefont {Szameit}},\ }\bibfield  {title} {\bibinfo {title}
  {Topological funneling of light},\ }\href
  {https://doi.org/10.1126/science.aaz8727} {\bibfield  {journal} {\bibinfo
  {journal} {Science}\ }\textbf {\bibinfo {volume} {368}},\ \bibinfo {pages}
  {311} (\bibinfo {year} {2020})}\BibitemShut {NoStop}%
\bibitem [{\citenamefont {Yokomizo}\ and\ \citenamefont
  {Murakami}(2019)}]{Yokomizo2019}%
  \BibitemOpen
  \bibfield  {author} {\bibinfo {author} {\bibfnamefont {K.}~\bibnamefont
  {Yokomizo}}\ and\ \bibinfo {author} {\bibfnamefont {S.}~\bibnamefont
  {Murakami}},\ }\bibfield  {title} {\bibinfo {title} {Non-bloch band theory of
  non-hermitian systems},\ }\href
  {https://doi.org/10.1103/PhysRevLett.123.066404} {\bibfield  {journal}
  {\bibinfo  {journal} {Phys. Rev. Lett.}\ }\textbf {\bibinfo {volume} {123}},\
  \bibinfo {pages} {066404} (\bibinfo {year} {2019})}\BibitemShut {NoStop}%
\bibitem [{\citenamefont {Zhang}\ \emph {et~al.}(2020)\citenamefont {Zhang},
  \citenamefont {Yang},\ and\ \citenamefont {Fang}}]{Zhang2020correspondence}%
  \BibitemOpen
  \bibfield  {author} {\bibinfo {author} {\bibfnamefont {K.}~\bibnamefont
  {Zhang}}, \bibinfo {author} {\bibfnamefont {Z.}~\bibnamefont {Yang}},\ and\
  \bibinfo {author} {\bibfnamefont {C.}~\bibnamefont {Fang}},\ }\bibfield
  {title} {\bibinfo {title} {Correspondence between winding numbers and skin
  modes in non-hermitian systems},\ }\href
  {https://doi.org/10.1103/PhysRevLett.125.126402} {\bibfield  {journal}
  {\bibinfo  {journal} {Phys. Rev. Lett.}\ }\textbf {\bibinfo {volume} {125}},\
  \bibinfo {pages} {126402} (\bibinfo {year} {2020})}\BibitemShut {NoStop}%
\bibitem [{\citenamefont {Hatano}\ and\ \citenamefont
  {Nelson}(1996)}]{Hatano1996}%
  \BibitemOpen
  \bibfield  {author} {\bibinfo {author} {\bibfnamefont {N.}~\bibnamefont
  {Hatano}}\ and\ \bibinfo {author} {\bibfnamefont {D.~R.}\ \bibnamefont
  {Nelson}},\ }\bibfield  {title} {\bibinfo {title} {Localization transitions
  in non-hermitian quantum mechanics},\ }\href
  {https://doi.org/10.1103/PhysRevLett.77.570} {\bibfield  {journal} {\bibinfo
  {journal} {Phys. Rev. Lett.}\ }\textbf {\bibinfo {volume} {77}},\ \bibinfo
  {pages} {570} (\bibinfo {year} {1996})}\BibitemShut {NoStop}%
\bibitem [{\citenamefont {Hatano}\ and\ \citenamefont
  {Nelson}(1997)}]{Hatano1997vortex}%
  \BibitemOpen
  \bibfield  {author} {\bibinfo {author} {\bibfnamefont {N.}~\bibnamefont
  {Hatano}}\ and\ \bibinfo {author} {\bibfnamefont {D.~R.}\ \bibnamefont
  {Nelson}},\ }\bibfield  {title} {\bibinfo {title} {Vortex pinning and
  non-hermitian quantum mechanics},\ }\href
  {https://doi.org/10.1103/PhysRevB.56.8651} {\bibfield  {journal} {\bibinfo
  {journal} {Phys. Rev. B}\ }\textbf {\bibinfo {volume} {56}},\ \bibinfo
  {pages} {8651} (\bibinfo {year} {1997})}\BibitemShut {NoStop}%
\bibitem [{\citenamefont {Hatano}\ and\ \citenamefont
  {Nelson}(1998)}]{Hatano1998non-Hermitian}%
  \BibitemOpen
  \bibfield  {author} {\bibinfo {author} {\bibfnamefont {N.}~\bibnamefont
  {Hatano}}\ and\ \bibinfo {author} {\bibfnamefont {D.~R.}\ \bibnamefont
  {Nelson}},\ }\bibfield  {title} {\bibinfo {title} {Non-hermitian
  delocalization and eigenfunctions},\ }\href
  {https://doi.org/10.1103/PhysRevB.58.8384} {\bibfield  {journal} {\bibinfo
  {journal} {Phys. Rev. B}\ }\textbf {\bibinfo {volume} {58}},\ \bibinfo
  {pages} {8384} (\bibinfo {year} {1998})}\BibitemShut {NoStop}%
\bibitem [{\citenamefont {Jiang}\ \emph {et~al.}(2019)\citenamefont {Jiang},
  \citenamefont {Lang}, \citenamefont {Yang}, \citenamefont {Zhu},\ and\
  \citenamefont {Chen}}]{jiang2019interplay}%
  \BibitemOpen
  \bibfield  {author} {\bibinfo {author} {\bibfnamefont {H.}~\bibnamefont
  {Jiang}}, \bibinfo {author} {\bibfnamefont {L.-J.}\ \bibnamefont {Lang}},
  \bibinfo {author} {\bibfnamefont {C.}~\bibnamefont {Yang}}, \bibinfo {author}
  {\bibfnamefont {S.-L.}\ \bibnamefont {Zhu}},\ and\ \bibinfo {author}
  {\bibfnamefont {S.}~\bibnamefont {Chen}},\ }\bibfield  {title} {\bibinfo
  {title} {Interplay of non-hermitian skin effects and anderson localization in
  nonreciprocal quasiperiodic lattices},\ }\href
  {https://doi.org/10.1103/PhysRevB.100.054301} {\bibfield  {journal} {\bibinfo
   {journal} {Phys. Rev. B}\ }\textbf {\bibinfo {volume} {100}},\ \bibinfo
  {pages} {054301} (\bibinfo {year} {2019})}\BibitemShut {NoStop}%
\bibitem [{\citenamefont
  {Longhi}(2019{\natexlab{a}})}]{longhi2019metal-insulator}%
  \BibitemOpen
  \bibfield  {author} {\bibinfo {author} {\bibfnamefont {S.}~\bibnamefont
  {Longhi}},\ }\bibfield  {title} {\bibinfo {title} {Metal-insulator phase
  transition in a non-hermitian aubry-andr\'e-harper model},\ }\href
  {https://doi.org/10.1103/PhysRevB.100.125157} {\bibfield  {journal} {\bibinfo
   {journal} {Phys. Rev. B}\ }\textbf {\bibinfo {volume} {100}},\ \bibinfo
  {pages} {125157} (\bibinfo {year} {2019}{\natexlab{a}})}\BibitemShut
  {NoStop}%
\bibitem [{\citenamefont {Liu}\ \emph {et~al.}(2020{\natexlab{a}})\citenamefont
  {Liu}, \citenamefont {Guo}, \citenamefont {Pu},\ and\ \citenamefont
  {Longhi}}]{liu2020generalized}%
  \BibitemOpen
  \bibfield  {author} {\bibinfo {author} {\bibfnamefont {T.}~\bibnamefont
  {Liu}}, \bibinfo {author} {\bibfnamefont {H.}~\bibnamefont {Guo}}, \bibinfo
  {author} {\bibfnamefont {Y.}~\bibnamefont {Pu}},\ and\ \bibinfo {author}
  {\bibfnamefont {S.}~\bibnamefont {Longhi}},\ }\bibfield  {title} {\bibinfo
  {title} {Generalized aubry-andr\'e self-duality and mobility edges in
  non-hermitian quasiperiodic lattices},\ }\href
  {https://doi.org/10.1103/PhysRevB.102.024205} {\bibfield  {journal} {\bibinfo
   {journal} {Phys. Rev. B}\ }\textbf {\bibinfo {volume} {102}},\ \bibinfo
  {pages} {024205} (\bibinfo {year} {2020}{\natexlab{a}})}\BibitemShut
  {NoStop}%
\bibitem [{\citenamefont {Liu}\ \emph {et~al.}(2020{\natexlab{b}})\citenamefont
  {Liu}, \citenamefont {Jiang}, \citenamefont {Cao},\ and\ \citenamefont
  {Chen}}]{liu2020non-Hermtian}%
  \BibitemOpen
  \bibfield  {author} {\bibinfo {author} {\bibfnamefont {Y.}~\bibnamefont
  {Liu}}, \bibinfo {author} {\bibfnamefont {X.-P.}\ \bibnamefont {Jiang}},
  \bibinfo {author} {\bibfnamefont {J.}~\bibnamefont {Cao}},\ and\ \bibinfo
  {author} {\bibfnamefont {S.}~\bibnamefont {Chen}},\ }\bibfield  {title}
  {\bibinfo {title} {Non-hermitian mobility edges in one-dimensional
  quasicrystals with parity-time symmetry},\ }\href
  {https://doi.org/10.1103/PhysRevB.101.174205} {\bibfield  {journal} {\bibinfo
   {journal} {Phys. Rev. B}\ }\textbf {\bibinfo {volume} {101}},\ \bibinfo
  {pages} {174205} (\bibinfo {year} {2020}{\natexlab{b}})}\BibitemShut
  {NoStop}%
\bibitem [{\citenamefont {Liu}\ \emph {et~al.}(2021{\natexlab{a}})\citenamefont
  {Liu}, \citenamefont {Wang}, \citenamefont {Liu}, \citenamefont {Zhou},\ and\
  \citenamefont {Chen}}]{liu2021exact_mobility_edges}%
  \BibitemOpen
  \bibfield  {author} {\bibinfo {author} {\bibfnamefont {Y.}~\bibnamefont
  {Liu}}, \bibinfo {author} {\bibfnamefont {Y.}~\bibnamefont {Wang}}, \bibinfo
  {author} {\bibfnamefont {X.-J.}\ \bibnamefont {Liu}}, \bibinfo {author}
  {\bibfnamefont {Q.}~\bibnamefont {Zhou}},\ and\ \bibinfo {author}
  {\bibfnamefont {S.}~\bibnamefont {Chen}},\ }\bibfield  {title} {\bibinfo
  {title} {Exact mobility edges, $\mathcal{PT}$-symmetry breaking, and skin
  effect in one-dimensional non-hermitian quasicrystals},\ }\href
  {https://doi.org/10.1103/PhysRevB.103.014203} {\bibfield  {journal} {\bibinfo
   {journal} {Phys. Rev. B}\ }\textbf {\bibinfo {volume} {103}},\ \bibinfo
  {pages} {014203} (\bibinfo {year} {2021}{\natexlab{a}})}\BibitemShut
  {NoStop}%
\bibitem [{\citenamefont
  {Longhi}(2021{\natexlab{a}})}]{longhi2021phase_transition}%
  \BibitemOpen
  \bibfield  {author} {\bibinfo {author} {\bibfnamefont {S.}~\bibnamefont
  {Longhi}},\ }\bibfield  {title} {\bibinfo {title} {Phase transitions in a
  non-hermitian aubry-andr\'e-harper model},\ }\href
  {https://doi.org/10.1103/PhysRevB.103.054203} {\bibfield  {journal} {\bibinfo
   {journal} {Phys. Rev. B}\ }\textbf {\bibinfo {volume} {103}},\ \bibinfo
  {pages} {054203} (\bibinfo {year} {2021}{\natexlab{a}})}\BibitemShut
  {NoStop}%
\bibitem [{\citenamefont {Lin}\ \emph {et~al.}(2022{\natexlab{a}})\citenamefont
  {Lin}, \citenamefont {Li}, \citenamefont {Xiao}, \citenamefont {Wang},
  \citenamefont {Yi},\ and\ \citenamefont {Xue}}]{lin2022topological}%
  \BibitemOpen
  \bibfield  {author} {\bibinfo {author} {\bibfnamefont {Q.}~\bibnamefont
  {Lin}}, \bibinfo {author} {\bibfnamefont {T.}~\bibnamefont {Li}}, \bibinfo
  {author} {\bibfnamefont {L.}~\bibnamefont {Xiao}}, \bibinfo {author}
  {\bibfnamefont {K.}~\bibnamefont {Wang}}, \bibinfo {author} {\bibfnamefont
  {W.}~\bibnamefont {Yi}},\ and\ \bibinfo {author} {\bibfnamefont
  {P.}~\bibnamefont {Xue}},\ }\bibfield  {title} {\bibinfo {title} {Topological
  phase transitions and mobility edges in non-hermitian quasicrystals},\ }\href
  {https://doi.org/10.1103/PhysRevLett.129.113601} {\bibfield  {journal}
  {\bibinfo  {journal} {Phys. Rev. Lett.}\ }\textbf {\bibinfo {volume} {129}},\
  \bibinfo {pages} {113601} (\bibinfo {year} {2022}{\natexlab{a}})}\BibitemShut
  {NoStop}%
\bibitem [{\citenamefont {Liu}\ \emph {et~al.}(2021{\natexlab{b}})\citenamefont
  {Liu}, \citenamefont {Zhou},\ and\ \citenamefont
  {Chen}}]{liu2021localization}%
  \BibitemOpen
  \bibfield  {author} {\bibinfo {author} {\bibfnamefont {Y.}~\bibnamefont
  {Liu}}, \bibinfo {author} {\bibfnamefont {Q.}~\bibnamefont {Zhou}},\ and\
  \bibinfo {author} {\bibfnamefont {S.}~\bibnamefont {Chen}},\ }\bibfield
  {title} {\bibinfo {title} {Localization transition, spectrum structure, and
  winding numbers for one-dimensional non-hermitian quasicrystals},\ }\href
  {https://doi.org/10.1103/PhysRevB.104.024201} {\bibfield  {journal} {\bibinfo
   {journal} {Phys. Rev. B}\ }\textbf {\bibinfo {volume} {104}},\ \bibinfo
  {pages} {024201} (\bibinfo {year} {2021}{\natexlab{b}})}\BibitemShut
  {NoStop}%
\bibitem [{\citenamefont {Luo}\ \emph {et~al.}(2021)\citenamefont {Luo},
  \citenamefont {Ohtsuki},\ and\ \citenamefont {Shindou}}]{luo2021transfer}%
  \BibitemOpen
  \bibfield  {author} {\bibinfo {author} {\bibfnamefont {X.}~\bibnamefont
  {Luo}}, \bibinfo {author} {\bibfnamefont {T.}~\bibnamefont {Ohtsuki}},\ and\
  \bibinfo {author} {\bibfnamefont {R.}~\bibnamefont {Shindou}},\ }\bibfield
  {title} {\bibinfo {title} {Transfer matrix study of the anderson transition
  in non-hermitian systems},\ }\href
  {https://doi.org/10.1103/PhysRevB.104.104203} {\bibfield  {journal} {\bibinfo
   {journal} {Phys. Rev. B}\ }\textbf {\bibinfo {volume} {104}},\ \bibinfo
  {pages} {104203} (\bibinfo {year} {2021})}\BibitemShut {NoStop}%
\bibitem [{\citenamefont {Luo}\ \emph {et~al.}(2022)\citenamefont {Luo},
  \citenamefont {Xiao}, \citenamefont {Kawabata}, \citenamefont {Ohtsuki},\
  and\ \citenamefont {Shindou}}]{luo2022unifying}%
  \BibitemOpen
  \bibfield  {author} {\bibinfo {author} {\bibfnamefont {X.}~\bibnamefont
  {Luo}}, \bibinfo {author} {\bibfnamefont {Z.}~\bibnamefont {Xiao}}, \bibinfo
  {author} {\bibfnamefont {K.}~\bibnamefont {Kawabata}}, \bibinfo {author}
  {\bibfnamefont {T.}~\bibnamefont {Ohtsuki}},\ and\ \bibinfo {author}
  {\bibfnamefont {R.}~\bibnamefont {Shindou}},\ }\bibfield  {title} {\bibinfo
  {title} {Unifying the anderson transitions in hermitian and non-hermitian
  systems},\ }\href {https://doi.org/10.1103/PhysRevResearch.4.L022035}
  {\bibfield  {journal} {\bibinfo  {journal} {Phys. Rev. Res.}\ }\textbf
  {\bibinfo {volume} {4}},\ \bibinfo {pages} {L022035} (\bibinfo {year}
  {2022})}\BibitemShut {NoStop}%
\bibitem [{\citenamefont {Kawabata}\ and\ \citenamefont
  {Ryu}(2021)}]{Kawabata2021nonunitary}%
  \BibitemOpen
  \bibfield  {author} {\bibinfo {author} {\bibfnamefont {K.}~\bibnamefont
  {Kawabata}}\ and\ \bibinfo {author} {\bibfnamefont {S.}~\bibnamefont {Ryu}},\
  }\bibfield  {title} {\bibinfo {title} {Nonunitary scaling theory of
  non-hermitian localization},\ }\href
  {https://doi.org/10.1103/PhysRevLett.126.166801} {\bibfield  {journal}
  {\bibinfo  {journal} {Phys. Rev. Lett.}\ }\textbf {\bibinfo {volume} {126}},\
  \bibinfo {pages} {166801} (\bibinfo {year} {2021})}\BibitemShut {NoStop}%
\bibitem [{\citenamefont {Longhi}(2021{\natexlab{b}})}]{longhi2021maryland}%
  \BibitemOpen
  \bibfield  {author} {\bibinfo {author} {\bibfnamefont {S.}~\bibnamefont
  {Longhi}},\ }\bibfield  {title} {\bibinfo {title} {Non-hermitian maryland
  model},\ }\href {https://doi.org/10.1103/PhysRevB.103.224206} {\bibfield
  {journal} {\bibinfo  {journal} {Phys. Rev. B}\ }\textbf {\bibinfo {volume}
  {103}},\ \bibinfo {pages} {224206} (\bibinfo {year}
  {2021}{\natexlab{b}})}\BibitemShut {NoStop}%
\bibitem [{\citenamefont {Longhi}(2021{\natexlab{c}})}]{longhi2021spectral}%
  \BibitemOpen
  \bibfield  {author} {\bibinfo {author} {\bibfnamefont {S.}~\bibnamefont
  {Longhi}},\ }\bibfield  {title} {\bibinfo {title} {Spectral deformations in
  non-hermitian lattices with disorder and skin effect: A solvable model},\
  }\href {https://doi.org/10.1103/PhysRevB.103.144202} {\bibfield  {journal}
  {\bibinfo  {journal} {Phys. Rev. B}\ }\textbf {\bibinfo {volume} {103}},\
  \bibinfo {pages} {144202} (\bibinfo {year} {2021}{\natexlab{c}})}\BibitemShut
  {NoStop}%
\bibitem [{\citenamefont {Li}\ \emph {et~al.}(2022)\citenamefont {Li},
  \citenamefont {Zhang},\ and\ \citenamefont {Yi}}]{li2022engineering}%
  \BibitemOpen
  \bibfield  {author} {\bibinfo {author} {\bibfnamefont {T.}~\bibnamefont
  {Li}}, \bibinfo {author} {\bibfnamefont {Y.-S.}\ \bibnamefont {Zhang}},\ and\
  \bibinfo {author} {\bibfnamefont {W.}~\bibnamefont {Yi}},\ }\bibfield
  {title} {\bibinfo {title} {Engineering dissipative quasicrystals},\ }\href
  {https://doi.org/10.1103/PhysRevB.105.125111} {\bibfield  {journal} {\bibinfo
   {journal} {Phys. Rev. B}\ }\textbf {\bibinfo {volume} {105}},\ \bibinfo
  {pages} {125111} (\bibinfo {year} {2022})}\BibitemShut {NoStop}%
\bibitem [{\citenamefont {Wang}\ \emph
  {et~al.}(2024{\natexlab{a}})\citenamefont {Wang}, \citenamefont {Cheng},
  \citenamefont {Zou}, \citenamefont {Ge}, \citenamefont {Zhao}, \citenamefont
  {Si}, \citenamefont {Yuan}, \citenamefont {Sun}, \citenamefont {Xue},\ and\
  \citenamefont {Zhang}}]{wang2024disorderinduced}%
  \BibitemOpen
  \bibfield  {author} {\bibinfo {author} {\bibfnamefont {B.-B.}\ \bibnamefont
  {Wang}}, \bibinfo {author} {\bibfnamefont {Z.}~\bibnamefont {Cheng}},
  \bibinfo {author} {\bibfnamefont {H.-Y.}\ \bibnamefont {Zou}}, \bibinfo
  {author} {\bibfnamefont {Y.}~\bibnamefont {Ge}}, \bibinfo {author}
  {\bibfnamefont {K.-Q.}\ \bibnamefont {Zhao}}, \bibinfo {author}
  {\bibfnamefont {Q.-R.}\ \bibnamefont {Si}}, \bibinfo {author} {\bibfnamefont
  {S.-Q.}\ \bibnamefont {Yuan}}, \bibinfo {author} {\bibfnamefont {H.-X.}\
  \bibnamefont {Sun}}, \bibinfo {author} {\bibfnamefont {H.}~\bibnamefont
  {Xue}},\ and\ \bibinfo {author} {\bibfnamefont {B.}~\bibnamefont {Zhang}},\
  }\href {https://arxiv.org/abs/2402.10989} {\bibinfo {title} {Disorder-induced
  acoustic non-hermitian skin effect}} (\bibinfo {year} {2024}{\natexlab{a}}),\
  \Eprint {https://arxiv.org/abs/2402.10989} {arXiv:2402.10989
  [physics.class-ph]} \BibitemShut {NoStop}%
\bibitem [{\citenamefont {Longhi}(2019{\natexlab{b}})}]{longhi2019topological}%
  \BibitemOpen
  \bibfield  {author} {\bibinfo {author} {\bibfnamefont {S.}~\bibnamefont
  {Longhi}},\ }\bibfield  {title} {\bibinfo {title} {Topological phase
  transition in non-hermitian quasicrystals},\ }\href
  {https://doi.org/10.1103/PhysRevLett.122.237601} {\bibfield  {journal}
  {\bibinfo  {journal} {Phys. Rev. Lett.}\ }\textbf {\bibinfo {volume} {122}},\
  \bibinfo {pages} {237601} (\bibinfo {year} {2019}{\natexlab{b}})}\BibitemShut
  {NoStop}%
\bibitem [{\citenamefont {Tang}\ \emph {et~al.}(2020)\citenamefont {Tang},
  \citenamefont {Zhang}, \citenamefont {Zhang},\ and\ \citenamefont
  {Zhang}}]{tang2020topological_anderson}%
  \BibitemOpen
  \bibfield  {author} {\bibinfo {author} {\bibfnamefont {L.-Z.}\ \bibnamefont
  {Tang}}, \bibinfo {author} {\bibfnamefont {L.-F.}\ \bibnamefont {Zhang}},
  \bibinfo {author} {\bibfnamefont {G.-Q.}\ \bibnamefont {Zhang}},\ and\
  \bibinfo {author} {\bibfnamefont {D.-W.}\ \bibnamefont {Zhang}},\ }\bibfield
  {title} {\bibinfo {title} {Topological anderson insulators in two-dimensional
  non-hermitian disordered systems},\ }\href
  {https://doi.org/10.1103/PhysRevA.101.063612} {\bibfield  {journal} {\bibinfo
   {journal} {Phys. Rev. A}\ }\textbf {\bibinfo {volume} {101}},\ \bibinfo
  {pages} {063612} (\bibinfo {year} {2020})}\BibitemShut {NoStop}%
\bibitem [{\citenamefont {Lin}\ \emph {et~al.}(2022{\natexlab{b}})\citenamefont
  {Lin}, \citenamefont {Li}, \citenamefont {Xiao}, \citenamefont {Wang},
  \citenamefont {Yi},\ and\ \citenamefont {Xue}}]{lin2022observation}%
  \BibitemOpen
  \bibfield  {author} {\bibinfo {author} {\bibfnamefont {Q.}~\bibnamefont
  {Lin}}, \bibinfo {author} {\bibfnamefont {T.}~\bibnamefont {Li}}, \bibinfo
  {author} {\bibfnamefont {L.}~\bibnamefont {Xiao}}, \bibinfo {author}
  {\bibfnamefont {K.}~\bibnamefont {Wang}}, \bibinfo {author} {\bibfnamefont
  {W.}~\bibnamefont {Yi}},\ and\ \bibinfo {author} {\bibfnamefont
  {P.}~\bibnamefont {Xue}},\ }\bibfield  {title} {\bibinfo {title} {Observation
  of non-hermitian topological anderson insulator in quantum dynamics},\
  }\href@noop {} {\bibfield  {journal} {\bibinfo  {journal} {Nature
  Communications}\ }\textbf {\bibinfo {volume} {13}},\ \bibinfo {pages} {3229}
  (\bibinfo {year} {2022}{\natexlab{b}})}\BibitemShut {NoStop}%
\bibitem [{\citenamefont {Cai}(2021)}]{cai2021localization}%
  \BibitemOpen
  \bibfield  {author} {\bibinfo {author} {\bibfnamefont {X.}~\bibnamefont
  {Cai}},\ }\bibfield  {title} {\bibinfo {title} {Localization and topological
  phase transitions in non-hermitian aubry-andr\'e-harper models with $p$-wave
  pairing},\ }\href {https://doi.org/10.1103/PhysRevB.103.214202} {\bibfield
  {journal} {\bibinfo  {journal} {Phys. Rev. B}\ }\textbf {\bibinfo {volume}
  {103}},\ \bibinfo {pages} {214202} (\bibinfo {year} {2021})}\BibitemShut
  {NoStop}%
\bibitem [{\citenamefont {Wang}\ \emph {et~al.}(2021)\citenamefont {Wang},
  \citenamefont {Li}, \citenamefont {Xiao}, \citenamefont {Han}, \citenamefont
  {Yi},\ and\ \citenamefont {Xue}}]{wang2021detecting}%
  \BibitemOpen
  \bibfield  {author} {\bibinfo {author} {\bibfnamefont {K.}~\bibnamefont
  {Wang}}, \bibinfo {author} {\bibfnamefont {T.}~\bibnamefont {Li}}, \bibinfo
  {author} {\bibfnamefont {L.}~\bibnamefont {Xiao}}, \bibinfo {author}
  {\bibfnamefont {Y.}~\bibnamefont {Han}}, \bibinfo {author} {\bibfnamefont
  {W.}~\bibnamefont {Yi}},\ and\ \bibinfo {author} {\bibfnamefont
  {P.}~\bibnamefont {Xue}},\ }\bibfield  {title} {\bibinfo {title} {Detecting
  non-bloch topological invariants in quantum dynamics},\ }\href
  {https://doi.org/10.1103/PhysRevLett.127.270602} {\bibfield  {journal}
  {\bibinfo  {journal} {Phys. Rev. Lett.}\ }\textbf {\bibinfo {volume} {127}},\
  \bibinfo {pages} {270602} (\bibinfo {year} {2021})}\BibitemShut {NoStop}%
\bibitem [{\citenamefont {Claes}\ and\ \citenamefont
  {Hughes}(2021)}]{claes2021skin}%
  \BibitemOpen
  \bibfield  {author} {\bibinfo {author} {\bibfnamefont {J.}~\bibnamefont
  {Claes}}\ and\ \bibinfo {author} {\bibfnamefont {T.~L.}\ \bibnamefont
  {Hughes}},\ }\bibfield  {title} {\bibinfo {title} {Skin effect and winding
  number in disordered non-hermitian systems},\ }\href
  {https://doi.org/10.1103/PhysRevB.103.L140201} {\bibfield  {journal}
  {\bibinfo  {journal} {Phys. Rev. B}\ }\textbf {\bibinfo {volume} {103}},\
  \bibinfo {pages} {L140201} (\bibinfo {year} {2021})}\BibitemShut {NoStop}%
\bibitem [{\citenamefont {Wang}\ \emph {et~al.}(2022)\citenamefont {Wang},
  \citenamefont {Wang},\ and\ \citenamefont {Ma}}]{wang2022non}%
  \BibitemOpen
  \bibfield  {author} {\bibinfo {author} {\bibfnamefont {W.}~\bibnamefont
  {Wang}}, \bibinfo {author} {\bibfnamefont {X.}~\bibnamefont {Wang}},\ and\
  \bibinfo {author} {\bibfnamefont {G.}~\bibnamefont {Ma}},\ }\bibfield
  {title} {\bibinfo {title} {Non-hermitian morphing of topological modes},\
  }\href {https://doi.org/10.1038/s41586-022-04929-1} {\bibfield  {journal}
  {\bibinfo  {journal} {Nature}\ }\textbf {\bibinfo {volume} {608}},\ \bibinfo
  {pages} {50–55} (\bibinfo {year} {2022})}\BibitemShut {NoStop}%
\bibitem [{\citenamefont {Slim}\ \emph {et~al.}(2024)\citenamefont {Slim},
  \citenamefont {Wanjura}, \citenamefont {Brunelli}, \citenamefont {del Pino},
  \citenamefont {Nunnenkamp},\ and\ \citenamefont {Verhagen}}]{Slim_2024}%
  \BibitemOpen
  \bibfield  {author} {\bibinfo {author} {\bibfnamefont {J.~J.}\ \bibnamefont
  {Slim}}, \bibinfo {author} {\bibfnamefont {C.~C.}\ \bibnamefont {Wanjura}},
  \bibinfo {author} {\bibfnamefont {M.}~\bibnamefont {Brunelli}}, \bibinfo
  {author} {\bibfnamefont {J.}~\bibnamefont {del Pino}}, \bibinfo {author}
  {\bibfnamefont {A.}~\bibnamefont {Nunnenkamp}},\ and\ \bibinfo {author}
  {\bibfnamefont {E.}~\bibnamefont {Verhagen}},\ }\bibfield  {title} {\bibinfo
  {title} {Optomechanical realization of the bosonic kitaev chain},\ }\href
  {https://doi.org/10.1038/s41586-024-07174-w} {\bibfield  {journal} {\bibinfo
  {journal} {Nature}\ }\textbf {\bibinfo {volume} {627}},\ \bibinfo {pages}
  {767–771} (\bibinfo {year} {2024})}\BibitemShut {NoStop}%
\bibitem [{\citenamefont {Wanjura}\ \emph {et~al.}(2020)\citenamefont
  {Wanjura}, \citenamefont {Brunelli},\ and\ \citenamefont
  {Nunnenkamp}}]{Wanjura2019}%
  \BibitemOpen
  \bibfield  {author} {\bibinfo {author} {\bibfnamefont {C.~C.}\ \bibnamefont
  {Wanjura}}, \bibinfo {author} {\bibfnamefont {M.}~\bibnamefont {Brunelli}},\
  and\ \bibinfo {author} {\bibfnamefont {A.}~\bibnamefont {Nunnenkamp}},\
  }\bibfield  {title} {\bibinfo {title} {Topological framework for directional
  amplification in driven-dissipative cavity arrays},\ }\href
  {https://doi.org/10.1038/s41467-020-16863-9} {\bibfield  {journal} {\bibinfo
  {journal} {Nature communications}\ }\textbf {\bibinfo {volume} {11}},\
  \bibinfo {pages} {3149} (\bibinfo {year} {2020})}\BibitemShut {NoStop}%
\bibitem [{\citenamefont {Xue}\ \emph {et~al.}(2021)\citenamefont {Xue},
  \citenamefont {Li}, \citenamefont {Hu}, \citenamefont {Song},\ and\
  \citenamefont {Wang}}]{xue2021simple}%
  \BibitemOpen
  \bibfield  {author} {\bibinfo {author} {\bibfnamefont {W.-T.}\ \bibnamefont
  {Xue}}, \bibinfo {author} {\bibfnamefont {M.-R.}\ \bibnamefont {Li}},
  \bibinfo {author} {\bibfnamefont {Y.-M.}\ \bibnamefont {Hu}}, \bibinfo
  {author} {\bibfnamefont {F.}~\bibnamefont {Song}},\ and\ \bibinfo {author}
  {\bibfnamefont {Z.}~\bibnamefont {Wang}},\ }\bibfield  {title} {\bibinfo
  {title} {Simple formulas of directional amplification from non-bloch band
  theory},\ }\href {https://doi.org/10.1103/PhysRevB.103.L241408} {\bibfield
  {journal} {\bibinfo  {journal} {Phys. Rev. B}\ }\textbf {\bibinfo {volume}
  {103}},\ \bibinfo {pages} {L241408} (\bibinfo {year} {2021})}\BibitemShut
  {NoStop}%
\bibitem [{\citenamefont {Borgnia}\ \emph {et~al.}(2020)\citenamefont
  {Borgnia}, \citenamefont {Kruchkov},\ and\ \citenamefont
  {Slager}}]{Borgnia2020nonhermitian}%
  \BibitemOpen
  \bibfield  {author} {\bibinfo {author} {\bibfnamefont {D.~S.}\ \bibnamefont
  {Borgnia}}, \bibinfo {author} {\bibfnamefont {A.~J.}\ \bibnamefont
  {Kruchkov}},\ and\ \bibinfo {author} {\bibfnamefont {R.-J.}\ \bibnamefont
  {Slager}},\ }\bibfield  {title} {\bibinfo {title} {Non-hermitian boundary
  modes and topology},\ }\href {https://doi.org/10.1103/PhysRevLett.124.056802}
  {\bibfield  {journal} {\bibinfo  {journal} {Phys. Rev. Lett.}\ }\textbf
  {\bibinfo {volume} {124}},\ \bibinfo {pages} {056802} (\bibinfo {year}
  {2020})}\BibitemShut {NoStop}%
\bibitem [{\citenamefont {Zirnstein}\ \emph {et~al.}(2021)\citenamefont
  {Zirnstein}, \citenamefont {Refael},\ and\ \citenamefont
  {Rosenow}}]{Zirnstein2021}%
  \BibitemOpen
  \bibfield  {author} {\bibinfo {author} {\bibfnamefont {H.-G.}\ \bibnamefont
  {Zirnstein}}, \bibinfo {author} {\bibfnamefont {G.}~\bibnamefont {Refael}},\
  and\ \bibinfo {author} {\bibfnamefont {B.}~\bibnamefont {Rosenow}},\
  }\bibfield  {title} {\bibinfo {title} {Bulk-boundary correspondence for
  non-hermitian hamiltonians via green functions},\ }\href
  {https://doi.org/10.1103/PhysRevLett.126.216407} {\bibfield  {journal}
  {\bibinfo  {journal} {Phys. Rev. Lett.}\ }\textbf {\bibinfo {volume} {126}},\
  \bibinfo {pages} {216407} (\bibinfo {year} {2021})}\BibitemShut {NoStop}%
\bibitem [{\citenamefont {Zirnstein}\ and\ \citenamefont
  {Rosenow}(2021)}]{Zirnstein2021exponential}%
  \BibitemOpen
  \bibfield  {author} {\bibinfo {author} {\bibfnamefont {H.-G.}\ \bibnamefont
  {Zirnstein}}\ and\ \bibinfo {author} {\bibfnamefont {B.}~\bibnamefont
  {Rosenow}},\ }\bibfield  {title} {\bibinfo {title} {Exponentially growing
  bulk green functions as signature of nontrivial non-hermitian winding number
  in one dimension},\ }\href {https://doi.org/10.1103/PhysRevB.103.195157}
  {\bibfield  {journal} {\bibinfo  {journal} {Phys. Rev. B}\ }\textbf {\bibinfo
  {volume} {103}},\ \bibinfo {pages} {195157} (\bibinfo {year}
  {2021})}\BibitemShut {NoStop}%
\bibitem [{\citenamefont {Hu}\ and\ \citenamefont {Wang}(2023)}]{hu2023green}%
  \BibitemOpen
  \bibfield  {author} {\bibinfo {author} {\bibfnamefont {Y.-M.}\ \bibnamefont
  {Hu}}\ and\ \bibinfo {author} {\bibfnamefont {Z.}~\bibnamefont {Wang}},\
  }\bibfield  {title} {\bibinfo {title} {Green's functions of multiband
  non-hermitian systems},\ }\href
  {https://doi.org/10.1103/PhysRevResearch.5.043073} {\bibfield  {journal}
  {\bibinfo  {journal} {Phys. Rev. Res.}\ }\textbf {\bibinfo {volume} {5}},\
  \bibinfo {pages} {043073} (\bibinfo {year} {2023})}\BibitemShut {NoStop}%
\bibitem [{\citenamefont {Wanjura}\ \emph {et~al.}(2021)\citenamefont
  {Wanjura}, \citenamefont {Brunelli},\ and\ \citenamefont
  {Nunnenkamp}}]{Wanjura2021disorder}%
  \BibitemOpen
  \bibfield  {author} {\bibinfo {author} {\bibfnamefont {C.~C.}\ \bibnamefont
  {Wanjura}}, \bibinfo {author} {\bibfnamefont {M.}~\bibnamefont {Brunelli}},\
  and\ \bibinfo {author} {\bibfnamefont {A.}~\bibnamefont {Nunnenkamp}},\
  }\bibfield  {title} {\bibinfo {title} {Correspondence between non-hermitian
  topology and directional amplification in the presence of disorder},\ }\href
  {https://doi.org/10.1103/PhysRevLett.127.213601} {\bibfield  {journal}
  {\bibinfo  {journal} {Phys. Rev. Lett.}\ }\textbf {\bibinfo {volume} {127}},\
  \bibinfo {pages} {213601} (\bibinfo {year} {2021})}\BibitemShut {NoStop}%
\bibitem [{\citenamefont {Trefethen}\ \emph {et~al.}(2001)\citenamefont
  {Trefethen}, \citenamefont {Contedini},\ and\ \citenamefont
  {Embree}}]{Trefethen2001spectra}%
  \BibitemOpen
  \bibfield  {author} {\bibinfo {author} {\bibfnamefont {L.~N.}\ \bibnamefont
  {Trefethen}}, \bibinfo {author} {\bibfnamefont {M.}~\bibnamefont
  {Contedini}},\ and\ \bibinfo {author} {\bibfnamefont {M.}~\bibnamefont
  {Embree}},\ }\bibfield  {title} {\bibinfo {title} {Spectra, pseudospectra,
  and localization for random bidiagonal matrices},\ }\href
  {https://doi.org/https://doi.org/10.1002/cpa.4} {\bibfield  {journal}
  {\bibinfo  {journal} {Communications on Pure and Applied Mathematics}\
  }\textbf {\bibinfo {volume} {54}},\ \bibinfo {pages} {595} (\bibinfo {year}
  {2001})}\BibitemShut {NoStop}%
\bibitem [{\citenamefont {Trefethen}\ and\ \citenamefont
  {Embree}(2020)}]{TrefethenEmbree+2020}%
  \BibitemOpen
  \bibfield  {author} {\bibinfo {author} {\bibfnamefont {L.~N.}\ \bibnamefont
  {Trefethen}}\ and\ \bibinfo {author} {\bibfnamefont {M.}~\bibnamefont
  {Embree}},\ }\href {https://doi.org/doi:10.1515/9780691213101} {\emph
  {\bibinfo {title} {Spectra and Pseudospectra: The Behavior of Nonnormal
  Matrices and Operators}}}\ (\bibinfo  {publisher} {Princeton University
  Press, Princeton},\ \bibinfo {year} {2020})\BibitemShut {NoStop}%
\bibitem [{\citenamefont {Touchette}(2009)}]{touchette2009large}%
  \BibitemOpen
  \bibfield  {author} {\bibinfo {author} {\bibfnamefont {H.}~\bibnamefont
  {Touchette}},\ }\bibfield  {title} {\bibinfo {title} {The large deviation
  approach to statistical mechanics},\ }\href
  {https://doi.org/https://doi.org/10.1016/j.physrep.2009.05.002} {\bibfield
  {journal} {\bibinfo  {journal} {Physics Reports}\ }\textbf {\bibinfo {volume}
  {478}},\ \bibinfo {pages} {1} (\bibinfo {year} {2009})}\BibitemShut {NoStop}%
\bibitem [{\citenamefont {Touchette}(2012)}]{touchette2012basic}%
  \BibitemOpen
  \bibfield  {author} {\bibinfo {author} {\bibfnamefont {H.}~\bibnamefont
  {Touchette}},\ }\href@noop {} {\bibinfo {title} {A basic introduction to
  large deviations: Theory, applications, simulations}} (\bibinfo {year}
  {2012}),\ \Eprint {https://arxiv.org/abs/1106.4146} {arXiv:1106.4146
  [cond-mat.stat-mech]} \BibitemShut {NoStop}%
\bibitem [{\citenamefont {Giardin\`a}\ \emph {et~al.}(2006)\citenamefont
  {Giardin\`a}, \citenamefont {Kurchan},\ and\ \citenamefont
  {Peliti}}]{Giardina2006direct}%
  \BibitemOpen
  \bibfield  {author} {\bibinfo {author} {\bibfnamefont {C.}~\bibnamefont
  {Giardin\`a}}, \bibinfo {author} {\bibfnamefont {J.}~\bibnamefont
  {Kurchan}},\ and\ \bibinfo {author} {\bibfnamefont {L.}~\bibnamefont
  {Peliti}},\ }\bibfield  {title} {\bibinfo {title} {Direct evaluation of
  large-deviation functions},\ }\href
  {https://doi.org/10.1103/PhysRevLett.96.120603} {\bibfield  {journal}
  {\bibinfo  {journal} {Phys. Rev. Lett.}\ }\textbf {\bibinfo {volume} {96}},\
  \bibinfo {pages} {120603} (\bibinfo {year} {2006})}\BibitemShut {NoStop}%
\bibitem [{\citenamefont {Garrahan}\ \emph {et~al.}(2007)\citenamefont
  {Garrahan}, \citenamefont {Jack}, \citenamefont {Lecomte}, \citenamefont
  {Pitard}, \citenamefont {van Duijvendijk},\ and\ \citenamefont {van
  Wijland}}]{Garrahan2007dynamical}%
  \BibitemOpen
  \bibfield  {author} {\bibinfo {author} {\bibfnamefont {J.~P.}\ \bibnamefont
  {Garrahan}}, \bibinfo {author} {\bibfnamefont {R.~L.}\ \bibnamefont {Jack}},
  \bibinfo {author} {\bibfnamefont {V.}~\bibnamefont {Lecomte}}, \bibinfo
  {author} {\bibfnamefont {E.}~\bibnamefont {Pitard}}, \bibinfo {author}
  {\bibfnamefont {K.}~\bibnamefont {van Duijvendijk}},\ and\ \bibinfo {author}
  {\bibfnamefont {F.}~\bibnamefont {van Wijland}},\ }\bibfield  {title}
  {\bibinfo {title} {Dynamical first-order phase transition in kinetically
  constrained models of glasses},\ }\href
  {https://doi.org/10.1103/PhysRevLett.98.195702} {\bibfield  {journal}
  {\bibinfo  {journal} {Phys. Rev. Lett.}\ }\textbf {\bibinfo {volume} {98}},\
  \bibinfo {pages} {195702} (\bibinfo {year} {2007})}\BibitemShut {NoStop}%
\bibitem [{\citenamefont {Garrahan}\ and\ \citenamefont
  {Lesanovsky}(2010)}]{Garrahan2010thermodynamics}%
  \BibitemOpen
  \bibfield  {author} {\bibinfo {author} {\bibfnamefont {J.~P.}\ \bibnamefont
  {Garrahan}}\ and\ \bibinfo {author} {\bibfnamefont {I.}~\bibnamefont
  {Lesanovsky}},\ }\bibfield  {title} {\bibinfo {title} {Thermodynamics of
  quantum jump trajectories},\ }\href
  {https://doi.org/10.1103/PhysRevLett.104.160601} {\bibfield  {journal}
  {\bibinfo  {journal} {Phys. Rev. Lett.}\ }\textbf {\bibinfo {volume} {104}},\
  \bibinfo {pages} {160601} (\bibinfo {year} {2010})}\BibitemShut {NoStop}%
\bibitem [{\citenamefont {Chetrite}\ and\ \citenamefont
  {Touchette}(2013)}]{Chetrite2013nonequilibrium}%
  \BibitemOpen
  \bibfield  {author} {\bibinfo {author} {\bibfnamefont {R.}~\bibnamefont
  {Chetrite}}\ and\ \bibinfo {author} {\bibfnamefont {H.}~\bibnamefont
  {Touchette}},\ }\bibfield  {title} {\bibinfo {title} {Nonequilibrium
  microcanonical and canonical ensembles and their equivalence},\ }\href
  {https://doi.org/10.1103/PhysRevLett.111.120601} {\bibfield  {journal}
  {\bibinfo  {journal} {Phys. Rev. Lett.}\ }\textbf {\bibinfo {volume} {111}},\
  \bibinfo {pages} {120601} (\bibinfo {year} {2013})}\BibitemShut {NoStop}%
\bibitem [{\citenamefont {\ifmmode \check{Z}\else
  \v{Z}\fi{}nidari\ifmmode~\check{c}\else
  \v{c}\fi{}}(2014)}]{Znidaric2014exact}%
  \BibitemOpen
  \bibfield  {author} {\bibinfo {author} {\bibfnamefont {M.}~\bibnamefont
  {\ifmmode \check{Z}\else \v{Z}\fi{}nidari\ifmmode~\check{c}\else
  \v{c}\fi{}}},\ }\bibfield  {title} {\bibinfo {title} {Exact large-deviation
  statistics for a nonequilibrium quantum spin chain},\ }\href
  {https://doi.org/10.1103/PhysRevLett.112.040602} {\bibfield  {journal}
  {\bibinfo  {journal} {Phys. Rev. Lett.}\ }\textbf {\bibinfo {volume} {112}},\
  \bibinfo {pages} {040602} (\bibinfo {year} {2014})}\BibitemShut {NoStop}%
\bibitem [{\citenamefont {Carollo}\ \emph {et~al.}(2019)\citenamefont
  {Carollo}, \citenamefont {Jack},\ and\ \citenamefont
  {Garrahan}}]{Carollo2019unraveling}%
  \BibitemOpen
  \bibfield  {author} {\bibinfo {author} {\bibfnamefont {F.}~\bibnamefont
  {Carollo}}, \bibinfo {author} {\bibfnamefont {R.~L.}\ \bibnamefont {Jack}},\
  and\ \bibinfo {author} {\bibfnamefont {J.~P.}\ \bibnamefont {Garrahan}},\
  }\bibfield  {title} {\bibinfo {title} {Unraveling the large deviation
  statistics of markovian open quantum systems},\ }\href
  {https://doi.org/10.1103/PhysRevLett.122.130605} {\bibfield  {journal}
  {\bibinfo  {journal} {Phys. Rev. Lett.}\ }\textbf {\bibinfo {volume} {122}},\
  \bibinfo {pages} {130605} (\bibinfo {year} {2019})}\BibitemShut {NoStop}%
\bibitem [{\citenamefont {Chetrite}\ and\ \citenamefont
  {Touchette}(2015)}]{chetrite2015nonequilibrium}%
  \BibitemOpen
  \bibfield  {author} {\bibinfo {author} {\bibfnamefont {R.}~\bibnamefont
  {Chetrite}}\ and\ \bibinfo {author} {\bibfnamefont {H.}~\bibnamefont
  {Touchette}},\ }\bibfield  {title} {\bibinfo {title} {Nonequilibrium markov
  processes conditioned on large deviations},\ }in\ \href
  {https://doi.org/https://doi.org/10.1007/s00023-014-0375-8} {\emph {\bibinfo
  {booktitle} {Annales Henri Poincar{\'e}}}},\ Vol.~\bibinfo {volume} {16}\
  (\bibinfo {organization} {Springer},\ \bibinfo {year} {2015})\ pp.\ \bibinfo
  {pages} {2005--0,57}\BibitemShut {NoStop}%
\bibitem [{\citenamefont {Jack}\ and\ \citenamefont
  {Sollich}(2010)}]{jack2010large}%
  \BibitemOpen
  \bibfield  {author} {\bibinfo {author} {\bibfnamefont {R.~L.}\ \bibnamefont
  {Jack}}\ and\ \bibinfo {author} {\bibfnamefont {P.}~\bibnamefont {Sollich}},\
  }\bibfield  {title} {\bibinfo {title} {Large deviations and ensembles of
  trajectories in stochastic models},\ }\href
  {https://doi.org/https://doi.org/10.1143/PTPS.184.304} {\bibfield  {journal}
  {\bibinfo  {journal} {Progress of Theoretical Physics Supplement}\ }\textbf
  {\bibinfo {volume} {184}},\ \bibinfo {pages} {304} (\bibinfo {year}
  {2010})}\BibitemShut {NoStop}%
\bibitem [{\citenamefont {Hedges}\ \emph {et~al.}(2009)\citenamefont {Hedges},
  \citenamefont {Jack}, \citenamefont {Garrahan},\ and\ \citenamefont
  {Chandler}}]{hedges2009dynamic}%
  \BibitemOpen
  \bibfield  {author} {\bibinfo {author} {\bibfnamefont {L.~O.}\ \bibnamefont
  {Hedges}}, \bibinfo {author} {\bibfnamefont {R.~L.}\ \bibnamefont {Jack}},
  \bibinfo {author} {\bibfnamefont {J.~P.}\ \bibnamefont {Garrahan}},\ and\
  \bibinfo {author} {\bibfnamefont {D.}~\bibnamefont {Chandler}},\ }\bibfield
  {title} {\bibinfo {title} {Dynamic order-disorder in atomistic models of
  structural glass formers},\ }\href@noop {} {\bibfield  {journal} {\bibinfo
  {journal} {Science}\ }\textbf {\bibinfo {volume} {323}},\ \bibinfo {pages}
  {1309} (\bibinfo {year} {2009})}\BibitemShut {NoStop}%
\bibitem [{sup()}]{supplemental}%
  \BibitemOpen
  \href@noop {} {}\bibinfo {note} {See Supplemental Material.}\BibitemShut
  {Stop}%
\bibitem [{\citenamefont {Kunst}\ and\ \citenamefont
  {Dwivedi}(2019)}]{kunst2018transfer}%
  \BibitemOpen
  \bibfield  {author} {\bibinfo {author} {\bibfnamefont {F.~K.}\ \bibnamefont
  {Kunst}}\ and\ \bibinfo {author} {\bibfnamefont {V.}~\bibnamefont
  {Dwivedi}},\ }\bibfield  {title} {\bibinfo {title} {Non-hermitian systems and
  topology: A transfer-matrix perspective},\ }\href
  {https://doi.org/10.1103/PhysRevB.99.245116} {\bibfield  {journal} {\bibinfo
  {journal} {Phys. Rev. B}\ }\textbf {\bibinfo {volume} {99}},\ \bibinfo
  {pages} {245116} (\bibinfo {year} {2019})}\BibitemShut {NoStop}%
\bibitem [{\citenamefont {Slevin}\ and\ \citenamefont
  {Ohtsuki}(2014)}]{slevin2014critical}%
  \BibitemOpen
  \bibfield  {author} {\bibinfo {author} {\bibfnamefont {K.}~\bibnamefont
  {Slevin}}\ and\ \bibinfo {author} {\bibfnamefont {T.}~\bibnamefont
  {Ohtsuki}},\ }\bibfield  {title} {\bibinfo {title} {Critical exponent for the
  anderson transition in the three-dimensional orthogonal universality class},\
  }\href {https://doi.org/10.1088/1367-2630/16/1/015012} {\bibfield  {journal}
  {\bibinfo  {journal} {New Journal of Physics}\ }\textbf {\bibinfo {volume}
  {16}},\ \bibinfo {pages} {015012} (\bibinfo {year} {2014})}\BibitemShut
  {NoStop}%
\bibitem [{\citenamefont {Song}\ \emph {et~al.}(2019)\citenamefont {Song},
  \citenamefont {Yao},\ and\ \citenamefont {Wang}}]{song2019chiral}%
  \BibitemOpen
  \bibfield  {author} {\bibinfo {author} {\bibfnamefont {F.}~\bibnamefont
  {Song}}, \bibinfo {author} {\bibfnamefont {S.}~\bibnamefont {Yao}},\ and\
  \bibinfo {author} {\bibfnamefont {Z.}~\bibnamefont {Wang}},\ }\bibfield
  {title} {\bibinfo {title} {Non-hermitian skin effect and chiral damping in
  open quantum systems},\ }\href
  {https://doi.org/10.1103/PhysRevLett.123.170401} {\bibfield  {journal}
  {\bibinfo  {journal} {Phys. Rev. Lett.}\ }\textbf {\bibinfo {volume} {123}},\
  \bibinfo {pages} {170401} (\bibinfo {year} {2019})}\BibitemShut {NoStop}%
\bibitem [{\citenamefont {McDonald}\ \emph {et~al.}(2022)\citenamefont
  {McDonald}, \citenamefont {Hanai},\ and\ \citenamefont
  {Clerk}}]{McDonald2022nonequilibrium}%
  \BibitemOpen
  \bibfield  {author} {\bibinfo {author} {\bibfnamefont {A.}~\bibnamefont
  {McDonald}}, \bibinfo {author} {\bibfnamefont {R.}~\bibnamefont {Hanai}},\
  and\ \bibinfo {author} {\bibfnamefont {A.~A.}\ \bibnamefont {Clerk}},\
  }\bibfield  {title} {\bibinfo {title} {Nonequilibrium stationary states of
  quantum non-hermitian lattice models},\ }\href
  {https://doi.org/10.1103/PhysRevB.105.064302} {\bibfield  {journal} {\bibinfo
   {journal} {Phys. Rev. B}\ }\textbf {\bibinfo {volume} {105}},\ \bibinfo
  {pages} {064302} (\bibinfo {year} {2022})}\BibitemShut {NoStop}%
\bibitem [{\citenamefont {Hu}\ \emph {et~al.}(2023)\citenamefont {Hu},
  \citenamefont {Xue}, \citenamefont {Song},\ and\ \citenamefont
  {Wang}}]{hu2023manybody}%
  \BibitemOpen
  \bibfield  {author} {\bibinfo {author} {\bibfnamefont {Y.-M.}\ \bibnamefont
  {Hu}}, \bibinfo {author} {\bibfnamefont {W.-T.}\ \bibnamefont {Xue}},
  \bibinfo {author} {\bibfnamefont {F.}~\bibnamefont {Song}},\ and\ \bibinfo
  {author} {\bibfnamefont {Z.}~\bibnamefont {Wang}},\ }\bibfield  {title}
  {\bibinfo {title} {Steady-state edge burst: From free-particle systems to
  interaction-induced phenomena},\ }\href
  {https://doi.org/10.1103/PhysRevB.108.235422} {\bibfield  {journal} {\bibinfo
   {journal} {Phys. Rev. B}\ }\textbf {\bibinfo {volume} {108}},\ \bibinfo
  {pages} {235422} (\bibinfo {year} {2023})}\BibitemShut {NoStop}%
\bibitem [{\citenamefont {Hamazaki}\ \emph {et~al.}(2019)\citenamefont
  {Hamazaki}, \citenamefont {Kawabata},\ and\ \citenamefont
  {Ueda}}]{Hamazaki2019nonhermitian}%
  \BibitemOpen
  \bibfield  {author} {\bibinfo {author} {\bibfnamefont {R.}~\bibnamefont
  {Hamazaki}}, \bibinfo {author} {\bibfnamefont {K.}~\bibnamefont {Kawabata}},\
  and\ \bibinfo {author} {\bibfnamefont {M.}~\bibnamefont {Ueda}},\ }\bibfield
  {title} {\bibinfo {title} {Non-hermitian many-body localization},\ }\href
  {https://doi.org/10.1103/PhysRevLett.123.090603} {\bibfield  {journal}
  {\bibinfo  {journal} {Phys. Rev. Lett.}\ }\textbf {\bibinfo {volume} {123}},\
  \bibinfo {pages} {090603} (\bibinfo {year} {2019})}\BibitemShut {NoStop}%
\bibitem [{\citenamefont {Zhai}\ \emph {et~al.}(2020)\citenamefont {Zhai},
  \citenamefont {Yin},\ and\ \citenamefont {Huang}}]{zhai2020many-body}%
  \BibitemOpen
  \bibfield  {author} {\bibinfo {author} {\bibfnamefont {L.-J.}\ \bibnamefont
  {Zhai}}, \bibinfo {author} {\bibfnamefont {S.}~\bibnamefont {Yin}},\ and\
  \bibinfo {author} {\bibfnamefont {G.-Y.}\ \bibnamefont {Huang}},\ }\bibfield
  {title} {\bibinfo {title} {Many-body localization in a non-hermitian
  quasiperiodic system},\ }\href {https://doi.org/10.1103/PhysRevB.102.064206}
  {\bibfield  {journal} {\bibinfo  {journal} {Phys. Rev. B}\ }\textbf {\bibinfo
  {volume} {102}},\ \bibinfo {pages} {064206} (\bibinfo {year}
  {2020})}\BibitemShut {NoStop}%
\bibitem [{\citenamefont {Tang}\ \emph {et~al.}(2021)\citenamefont {Tang},
  \citenamefont {Zhang}, \citenamefont {Zhang},\ and\ \citenamefont
  {Zhang}}]{Tang2021Localization}%
  \BibitemOpen
  \bibfield  {author} {\bibinfo {author} {\bibfnamefont {L.-Z.}\ \bibnamefont
  {Tang}}, \bibinfo {author} {\bibfnamefont {G.-Q.}\ \bibnamefont {Zhang}},
  \bibinfo {author} {\bibfnamefont {L.-F.}\ \bibnamefont {Zhang}},\ and\
  \bibinfo {author} {\bibfnamefont {D.-W.}\ \bibnamefont {Zhang}},\ }\bibfield
  {title} {\bibinfo {title} {Localization and topological transitions in
  non-hermitian quasiperiodic lattices},\ }\href
  {https://doi.org/10.1103/PhysRevA.103.033325} {\bibfield  {journal} {\bibinfo
   {journal} {Phys. Rev. A}\ }\textbf {\bibinfo {volume} {103}},\ \bibinfo
  {pages} {033325} (\bibinfo {year} {2021})}\BibitemShut {NoStop}%
\bibitem [{\citenamefont {Heu\ss{}en}\ \emph {et~al.}(2021)\citenamefont
  {Heu\ss{}en}, \citenamefont {White},\ and\ \citenamefont
  {Refael}}]{Heuen2021extracting}%
  \BibitemOpen
  \bibfield  {author} {\bibinfo {author} {\bibfnamefont {S.}~\bibnamefont
  {Heu\ss{}en}}, \bibinfo {author} {\bibfnamefont {C.~D.}\ \bibnamefont
  {White}},\ and\ \bibinfo {author} {\bibfnamefont {G.}~\bibnamefont
  {Refael}},\ }\bibfield  {title} {\bibinfo {title} {Extracting many-body
  localization lengths with an imaginary vector potential},\ }\href
  {https://doi.org/10.1103/PhysRevB.103.064201} {\bibfield  {journal} {\bibinfo
   {journal} {Phys. Rev. B}\ }\textbf {\bibinfo {volume} {103}},\ \bibinfo
  {pages} {064201} (\bibinfo {year} {2021})}\BibitemShut {NoStop}%
\bibitem [{\citenamefont {Suthar}\ \emph {et~al.}(2022)\citenamefont {Suthar},
  \citenamefont {Wang}, \citenamefont {Huang}, \citenamefont {Jen},\ and\
  \citenamefont {You}}]{Suthar2022non-Hermitian}%
  \BibitemOpen
  \bibfield  {author} {\bibinfo {author} {\bibfnamefont {K.}~\bibnamefont
  {Suthar}}, \bibinfo {author} {\bibfnamefont {Y.-C.}\ \bibnamefont {Wang}},
  \bibinfo {author} {\bibfnamefont {Y.-P.}\ \bibnamefont {Huang}}, \bibinfo
  {author} {\bibfnamefont {H.~H.}\ \bibnamefont {Jen}},\ and\ \bibinfo {author}
  {\bibfnamefont {J.-S.}\ \bibnamefont {You}},\ }\bibfield  {title} {\bibinfo
  {title} {Non-hermitian many-body localization with open boundaries},\ }\href
  {https://doi.org/10.1103/PhysRevB.106.064208} {\bibfield  {journal} {\bibinfo
   {journal} {Phys. Rev. B}\ }\textbf {\bibinfo {volume} {106}},\ \bibinfo
  {pages} {064208} (\bibinfo {year} {2022})}\BibitemShut {NoStop}%
\bibitem [{\citenamefont {Wang}\ \emph {et~al.}(2023)\citenamefont {Wang},
  \citenamefont {Suthar}, \citenamefont {Jen}, \citenamefont {Hsu},\ and\
  \citenamefont {You}}]{Wang2023nonhemitian}%
  \BibitemOpen
  \bibfield  {author} {\bibinfo {author} {\bibfnamefont {Y.-C.}\ \bibnamefont
  {Wang}}, \bibinfo {author} {\bibfnamefont {K.}~\bibnamefont {Suthar}},
  \bibinfo {author} {\bibfnamefont {H.~H.}\ \bibnamefont {Jen}}, \bibinfo
  {author} {\bibfnamefont {Y.-T.}\ \bibnamefont {Hsu}},\ and\ \bibinfo {author}
  {\bibfnamefont {J.-S.}\ \bibnamefont {You}},\ }\bibfield  {title} {\bibinfo
  {title} {Non-hermitian skin effects on thermal and many-body localized
  phases},\ }\href {https://doi.org/10.1103/PhysRevB.107.L220205} {\bibfield
  {journal} {\bibinfo  {journal} {Phys. Rev. B}\ }\textbf {\bibinfo {volume}
  {107}},\ \bibinfo {pages} {L220205} (\bibinfo {year} {2023})}\BibitemShut
  {NoStop}%
\bibitem [{\citenamefont {O'Brien}\ and\ \citenamefont
  {Refael}(2023)}]{obrien2023probing}%
  \BibitemOpen
  \bibfield  {author} {\bibinfo {author} {\bibfnamefont {L.}~\bibnamefont
  {O'Brien}}\ and\ \bibinfo {author} {\bibfnamefont {G.}~\bibnamefont
  {Refael}},\ }\bibfield  {title} {\bibinfo {title} {Probing localization
  properties of many-body hamiltonians via an imaginary vector potential},\
  }\href {https://doi.org/10.1103/PhysRevB.108.184207} {\bibfield  {journal}
  {\bibinfo  {journal} {Phys. Rev. B}\ }\textbf {\bibinfo {volume} {108}},\
  \bibinfo {pages} {184207} (\bibinfo {year} {2023})}\BibitemShut {NoStop}%
\bibitem [{\citenamefont {Roccati}\ \emph {et~al.}(2024)\citenamefont
  {Roccati}, \citenamefont {Balducci}, \citenamefont {Shir},\ and\
  \citenamefont {Chenu}}]{roccati2024diagnosing}%
  \BibitemOpen
  \bibfield  {author} {\bibinfo {author} {\bibfnamefont {F.}~\bibnamefont
  {Roccati}}, \bibinfo {author} {\bibfnamefont {F.}~\bibnamefont {Balducci}},
  \bibinfo {author} {\bibfnamefont {R.}~\bibnamefont {Shir}},\ and\ \bibinfo
  {author} {\bibfnamefont {A.}~\bibnamefont {Chenu}},\ }\bibfield  {title}
  {\bibinfo {title} {Diagnosing non-hermitian many-body localization and
  quantum chaos via singular value decomposition},\ }\href
  {https://doi.org/10.1103/PhysRevB.109.L140201} {\bibfield  {journal}
  {\bibinfo  {journal} {Phys. Rev. B}\ }\textbf {\bibinfo {volume} {109}},\
  \bibinfo {pages} {L140201} (\bibinfo {year} {2024})}\BibitemShut {NoStop}%
\bibitem [{\citenamefont {Zhang}\ \emph
  {et~al.}(2022{\natexlab{b}})\citenamefont {Zhang}, \citenamefont {Yang},\
  and\ \citenamefont {Fang}}]{zhang2022universal}%
  \BibitemOpen
  \bibfield  {author} {\bibinfo {author} {\bibfnamefont {K.}~\bibnamefont
  {Zhang}}, \bibinfo {author} {\bibfnamefont {Z.}~\bibnamefont {Yang}},\ and\
  \bibinfo {author} {\bibfnamefont {C.}~\bibnamefont {Fang}},\ }\bibfield
  {title} {\bibinfo {title} {Universal non-hermitian skin effect in two and
  higher dimensions},\ }\href@noop {} {\bibfield  {journal} {\bibinfo
  {journal} {Nature communications}\ }\textbf {\bibinfo {volume} {13}},\
  \bibinfo {pages} {2496} (\bibinfo {year} {2022}{\natexlab{b}})}\BibitemShut
  {NoStop}%
\bibitem [{\citenamefont {Jiang}\ and\ \citenamefont
  {Lee}(2023)}]{Jiang2023Dimensional}%
  \BibitemOpen
  \bibfield  {author} {\bibinfo {author} {\bibfnamefont {H.}~\bibnamefont
  {Jiang}}\ and\ \bibinfo {author} {\bibfnamefont {C.~H.}\ \bibnamefont
  {Lee}},\ }\bibfield  {title} {\bibinfo {title} {Dimensional transmutation
  from non-hermiticity},\ }\href
  {https://doi.org/10.1103/PhysRevLett.131.076401} {\bibfield  {journal}
  {\bibinfo  {journal} {Phys. Rev. Lett.}\ }\textbf {\bibinfo {volume} {131}},\
  \bibinfo {pages} {076401} (\bibinfo {year} {2023})}\BibitemShut {NoStop}%
\bibitem [{\citenamefont {Wang}\ \emph
  {et~al.}(2024{\natexlab{b}})\citenamefont {Wang}, \citenamefont {Song},\ and\
  \citenamefont {Wang}}]{wang2022amoeba}%
  \BibitemOpen
  \bibfield  {author} {\bibinfo {author} {\bibfnamefont {H.-Y.}\ \bibnamefont
  {Wang}}, \bibinfo {author} {\bibfnamefont {F.}~\bibnamefont {Song}},\ and\
  \bibinfo {author} {\bibfnamefont {Z.}~\bibnamefont {Wang}},\ }\bibfield
  {title} {\bibinfo {title} {Amoeba formulation of non-bloch band theory in
  arbitrary dimensions},\ }\href {https://doi.org/10.1103/PhysRevX.14.021011}
  {\bibfield  {journal} {\bibinfo  {journal} {Phys. Rev. X}\ }\textbf {\bibinfo
  {volume} {14}},\ \bibinfo {pages} {021011} (\bibinfo {year}
  {2024}{\natexlab{b}})}\BibitemShut {NoStop}%
\end{thebibliography}%


\begin{thebibliography}{3}%
\makeatletter
\providecommand \@ifxundefined [1]{%
 \@ifx{#1\undefined}
}%
\providecommand \@ifnum [1]{%
 \ifnum #1\expandafter \@firstoftwo
 \else \expandafter \@secondoftwo
 \fi
}%
\providecommand \@ifx [1]{%
 \ifx #1\expandafter \@firstoftwo
 \else \expandafter \@secondoftwo
 \fi
}%
\providecommand \natexlab [1]{#1}%
\providecommand \enquote  [1]{``#1''}%
\providecommand \bibnamefont  [1]{#1}%
\providecommand \bibfnamefont [1]{#1}%
\providecommand \citenamefont [1]{#1}%
\providecommand \href@noop [0]{\@secondoftwo}%
\providecommand \href [0]{\begingroup \@sanitize@url \@href}%
\providecommand \@href[1]{\@@startlink{#1}\@@href}%
\providecommand \@@href[1]{\endgroup#1\@@endlink}%
\providecommand \@sanitize@url [0]{\catcode `\\12\catcode `\$12\catcode
  `\&12\catcode `\#12\catcode `\^12\catcode `\_12\catcode `\%12\relax}%
\providecommand \@@startlink[1]{}%
\providecommand \@@endlink[0]{}%
\providecommand \url  [0]{\begingroup\@sanitize@url \@url }%
\providecommand \@url [1]{\endgroup\@href {#1}{\urlprefix }}%
\providecommand \urlprefix  [0]{URL }%
\providecommand \Eprint [0]{\href }%
\providecommand \doibase [0]{https://doi.org/}%
\providecommand \selectlanguage [0]{\@gobble}%
\providecommand \bibinfo  [0]{\@secondoftwo}%
\providecommand \bibfield  [0]{\@secondoftwo}%
\providecommand \translation [1]{[#1]}%
\providecommand \BibitemOpen [0]{}%
\providecommand \bibitemStop [0]{}%
\providecommand \bibitemNoStop [0]{.\EOS\space}%
\providecommand \EOS [0]{\spacefactor3000\relax}%
\providecommand \BibitemShut  [1]{\csname bibitem#1\endcsname}%
\let\auto@bib@innerbib\@empty
\bibitem [{\citenamefont {Slevin}\ and\ \citenamefont
  {Ohtsuki}(2014)}]{slevin2014critical}%
  \BibitemOpen
  \bibfield  {author} {\bibinfo {author} {\bibfnamefont {K.}~\bibnamefont
  {Slevin}}\ and\ \bibinfo {author} {\bibfnamefont {T.}~\bibnamefont
  {Ohtsuki}},\ }\bibfield  {title} {\bibinfo {title} {Critical exponent for the
  anderson transition in the three-dimensional orthogonal universality class},\
  }\href {https://doi.org/10.1088/1367-2630/16/1/015012} {\bibfield  {journal}
  {\bibinfo  {journal} {New Journal of Physics}\ }\textbf {\bibinfo {volume}
  {16}},\ \bibinfo {pages} {015012} (\bibinfo {year} {2014})}\BibitemShut
  {NoStop}%
\bibitem [{\citenamefont {Luo}\ \emph {et~al.}(2021)\citenamefont {Luo},
  \citenamefont {Ohtsuki},\ and\ \citenamefont {Shindou}}]{luo2021transfer}%
  \BibitemOpen
  \bibfield  {author} {\bibinfo {author} {\bibfnamefont {X.}~\bibnamefont
  {Luo}}, \bibinfo {author} {\bibfnamefont {T.}~\bibnamefont {Ohtsuki}},\ and\
  \bibinfo {author} {\bibfnamefont {R.}~\bibnamefont {Shindou}},\ }\bibfield
  {title} {\bibinfo {title} {Transfer matrix study of the anderson transition
  in non-hermitian systems},\ }\href
  {https://doi.org/10.1103/PhysRevB.104.104203} {\bibfield  {journal} {\bibinfo
   {journal} {Phys. Rev. B}\ }\textbf {\bibinfo {volume} {104}},\ \bibinfo
  {pages} {104203} (\bibinfo {year} {2021})}\BibitemShut {NoStop}%
\bibitem [{\citenamefont {Xue}\ \emph {et~al.}(2021)\citenamefont {Xue},
  \citenamefont {Li}, \citenamefont {Hu}, \citenamefont {Song},\ and\
  \citenamefont {Wang}}]{xue2021simple}%
  \BibitemOpen
  \bibfield  {author} {\bibinfo {author} {\bibfnamefont {W.-T.}\ \bibnamefont
  {Xue}}, \bibinfo {author} {\bibfnamefont {M.-R.}\ \bibnamefont {Li}},
  \bibinfo {author} {\bibfnamefont {Y.-M.}\ \bibnamefont {Hu}}, \bibinfo
  {author} {\bibfnamefont {F.}~\bibnamefont {Song}},\ and\ \bibinfo {author}
  {\bibfnamefont {Z.}~\bibnamefont {Wang}},\ }\bibfield  {title} {\bibinfo
  {title} {Simple formulas of directional amplification from non-bloch band
  theory},\ }\href {https://doi.org/10.1103/PhysRevB.103.L241408} {\bibfield
  {journal} {\bibinfo  {journal} {Phys. Rev. B}\ }\textbf {\bibinfo {volume}
  {103}},\ \bibinfo {pages} {L241408} (\bibinfo {year} {2021})}\BibitemShut
  {NoStop}%
\end{thebibliography}%
\end{document}


\title{Supplemental material: Universal scaling of Green's functions in disordered non-Hermitian systems}	  
\author{Yin-Quan Huang}
\altaffiliation{These authors contributed equally to this work.}
\affiliation{ Institute for Advanced Study, Tsinghua University, Beijing,  100084, China }

\author{Yu-Min Hu}
\altaffiliation{These authors contributed equally to this work.}
\affiliation{ Institute for Advanced Study, Tsinghua University, Beijing,  100084, China }

\author{Wen-Tan Xue}
\affiliation{ Institute for Advanced Study, Tsinghua University, Beijing,  100084, China }
\affiliation{ Department of Physics, National University of Singapore, Singapore 117542, Singapore }

\author{Zhong Wang}
\altaffiliation{ wangzhongemail@tsinghua.edu.cn }
\affiliation{ Institute for Advanced Study, Tsinghua University, Beijing,  100084, China }	
	\maketitle
	
\section{The scaling factor of $G_m(E)$ in the algebraic region}\label{appendix:tc}
In the main text, we have concluded that the scaling behavior of $G_m(E)$ in the algebraic region is given by $G_m(E)\sim L^\alpha$ with  $\alpha= \sup_s\left[\frac{s}{I(s)}\right]$. $I(s)$ is the Fenchel-Legendre transform of $\Lambda(r)$, i.e., $I(s)=\sup_r[sr-\Lambda(r)]$. In this part, we will derive the scaling factor $\alpha$ for the algebraic behavior of $G_m(E)$. Under this circumstance, we always have $\Lambda(1)<0<\Lambda(+\infty)$, which is indicated by the relation $T_{\text{ave}}<1<T_{\max}$ presented in the main text.

We begin by assuming that the function $rs-\Lambda(r)$ takes the maximum at $r=r_m$. Then we have
\begin{equation}
	\left.\frac{\mathrm{d}}{\mathrm{d} r}[sr-\Lambda(r)]\right|_{r=r_m} =s-\Lambda'(r_m)=0,\label{apeq:zero1}
\end{equation}
where we have defined $\Lambda'(r)=\frac{\mathrm{d}}{\mathrm{d}r}\Lambda(r)$. Eq. \eqref{apeq:zero1} indicates that $r_m(s)$ is a function of $s$. We thus obtain
\begin{equation}
I(s)=sr_m(s)-\Lambda[r_m(s)].\label{apeq:zero2}
\end{equation}
We also assume that $\frac{s}{I(s)}$ takes the maximum at $s=s_0$. Then we get
\begin{equation}\label{apeq:zero3}
\frac{\mathrm{d}}{\mathrm{d} s}\left[\frac{s}{I(s)}\right]\Bigg|_{s=s_0}=\frac{1}{I(s_0)^2}[I(s_0)-s_0I'(s_0)]=0,
\end{equation}
where $I'(s)\equiv\frac{\mathrm{d}}{\mathrm{d}s}I(s)$. Now we differentiate Eq. \eqref{apeq:zero2} with respect to $s$ to get
\begin{equation}\label{apeq:zero4}
I'(s)=  r_m(s)+\left(s-\Lambda'[r_m(s)]\right)\frac{\mathrm{d} r_m(s)}{\mathrm{d}s}=r_m(s),
\end{equation}
where we have used Eq. \eqref{apeq:zero1}. When $s=s_0$, the combination of Eq.  \eqref{apeq:zero3} and Eq. \eqref{apeq:zero4} gives rise to
\begin{equation}\label{apeq:zero6}
	\alpha=\frac{s_0}{I(s_0)}=[I'(s_0)]^{-1}=[r_m(s_0)]^{-1}.
\end{equation}
Finally, when $s=s_0$, substituting Eq. \eqref{apeq:zero6} into Eq. \eqref{apeq:zero2} leads to
\begin{equation}
    \Lambda[r_m(s_0)]=0. 
\end{equation}
Therefore, the scaling behavior of $G_m(E)$ in the algebraic region is given by $G_m(E)\sim L^{1/r_0}$, where $r_0$ satisfies $\Lambda(r_0)=0$.

\section{The logarithmic scaling of the length of the possible largest $|S_n|$}
\begin{figure}
    \centering
    \includegraphics[width=8cm]{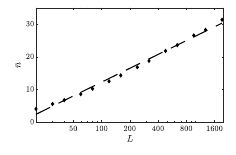}
    \caption{The logarithmic scaling of $n$ for the largest possible $|S_n|$. The model is $H_1$ in the main text. In the numerical calculation, we set $t=1.8$ and $E=1.5+1.2i$. $V_i$ is taken from a uniform distribution in the interval $[-1,1]$. The result is obtained by averaging over $10^4$ disorder realizations.}
    \label{sfig:logL}
\end{figure}

In the main text, we find that, in the algebraic scaling phase, the length $n$ of the possible largest $|S_n|$ exhibits a logarithmic scaling behavior with respect to the system size $L$. Here, we present additional numerical evidence to support this conclusion. Specifically, we first calculate the Green's function $G(E)=(E-H_1)^{-1}$ for the Hamiltonian $H_1$ in the main text. For each realization of the random potential in a finite system, we define $n\equiv |k-l|$ where $k$ and $l$ satisfy $|G_{kl}(E)|=\max_{ij}|G_{ij}(E)|$. Namely, $n$ provides the length of the largest possible $|S_n|$ in this finite system. Then we calculate the mean value of $n$ for different realizations of the disorder. Notably, the resulting mean value $\overline n$ depends on the system size $L$ in a logarithmic way, which is clearly evident in Fig. \ref{sfig:logL}. This result is consistent with the large deviation prediction in the main text.

\section{The Lyapunov exponents of transfer matrices}
In this section, we present a detailed discussion on the Lyapunov exponents (LEs) in general disordered non-Hermitian systems. We take the Hamiltonian $H_2$ in the main text as an example. All the results can be easily extended into general disordered non-Hermitian systems with finite hopping ranges. 

The central ingredients in calculating LEs are the transfer matrices generated from a certain random distribution. As shown in Eq.  (8) of the main text, the transfer matrices (to the right direction) of the Hamiltonian $H_2$ are given by 
\begin{equation}\label{apeq:T_matrix}
	T_i(E)=\begin{pmatrix}
		-\frac{t_{1}}{t_2} & \frac{E-V_i-t_0}{t_2}& -\frac{t_{-1}}{t_2}  &-\frac{t_{-2}}{t_2}\\
		 1 & 0 & 0 & 0\\
		 0 & 1 & 0 & 0\\
	 	 0 & 0 & 1 & 0
	\end{pmatrix},
\end{equation}
where $V_i$ is taken from a certain random distribution. 

In this section, we demonstrate how to extract LEs $\lambda_k$ from the random sequences $\mathcal{S}_{k,n}$ generated by QR decomposition \cite{slevin2014critical,luo2021transfer}.  Then we numerically show that certain LEs are related to the exponential factors of off-diagonal elements of non-Hermitian disordered Green's functions. Furthermore, we develop the method to calculate the cumulant generating functions in generic disordered non-Hermitian systems. 
\subsection{The calculation of generalized Lyapunov exponents}
\begin{figure}
    \centering
    \includegraphics{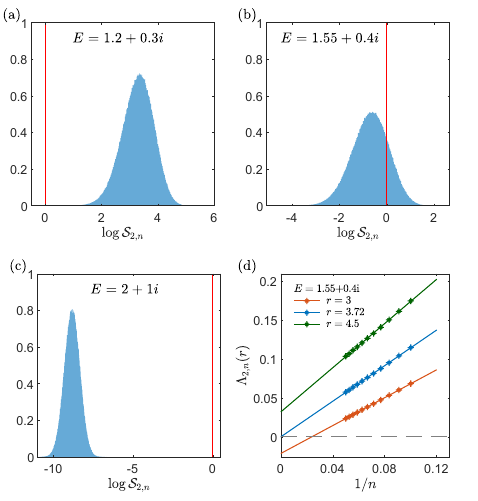}
    \caption{(a)-(c) The probability distributions of $\log\mathcal{S}_{2,n}$ obtained from $10^{6}$ samples, with the corresponding energies $E$ labeled in the plots. The parameters are $t_{-2}=t_2=0.1,\ t_{-1}=1.2,\ t_1=0.3,\ t_0=0.0$,  and $n=20$. The random potential $V$ is taken from the uniform distribution in the interval $[-0.6,0.6]$. Those positive $\log\mathcal{S}_{2,n}$ on the right of the red line imply possible amplification to the right direction. (d) Linear fittings of $\Lambda_{2,n}(r)=\frac{a}{n}+b$ for several $r$. When $n$ goes to infinity, $\Lambda_2(r)$ is estimated by the intercept $b$. When $E=1.55+0.4i$, the intercept approaches zero at $r=3.72$, namely, $\Lambda_2(3.72)\approx0$. Hence, we have $G_m(E)\sim L^{1/3.72}$ for $E=1.55+0.4i$ in the algebraic phase. }
    \label{sfig:distribution}
\end{figure}

In this part, we utilize the QR decomposition to show the numerical method to calculate the LEs $\lambda_k$ for $k=1,2,3,4$.  We start with the $n$-product $\mathcal{T}_n=\prod_{i=1}^{n}T_i=T_n\cdots T_1$ of the random transfer matrices $T_i$ in Eq. \eqref{apeq:T_matrix}. The QR decomposition of $\mathcal{T}_n=\prod_{i=1}^{n}T_i$ is
\begin{equation}\label{apeq:qr1}
	Q_n\mathcal{R}_n=\mathcal{T}_nQ_0
\end{equation}
where $Q_n$ is a unitary matrix and $\mathcal{R}_n$ is an upper triangular matrix. Additionally, $Q_0$ is an initial unitary matrix obtained by QR decomposition of $m$ multiplications of the random transfer matrix:
\begin{equation}\label{apeq:starting_matrix}
	Q_0R_0=\prod_{i=1}^{m}T^{(0)}_i.
\end{equation}
Here, the set of $T^{(0)}_i$ is another random sequence of the transfer matrix Eq. \eqref{apeq:T_matrix}. We set $m=100$ in our numerical simulation. The diagonal elements of $\mathcal{R}_n$ provide the LEs:
\begin{equation}\label{apeq:lypunov}
	\lambda_k=\lim\limits_{n\rightarrow\infty}\frac{1}{n}\log |(\mathcal{R}_n)_{k,k}|,
\end{equation}
which is a constant independent of different realizations of $\mathcal{T}_n$.

In the practical simulation where $n$ is large but finite, we define several random sequences as
\begin{equation}
    \mathcal{S}_{k,n}\equiv|(\mathcal{R}_n)_{k,k}|.
\end{equation}
 LEs are then obtained by $\lambda_k=\lim\limits_{n\to\infty}\frac{1}{n}\log\mathcal{S}_{k,n}$. To mitigate numerical errors, we extract $\mathcal{S}_{k,n}$ in the following process. Instead of directly implementing Eq. \eqref{apeq:qr1}, we first start with the initial unitary matrix $Q_0$ obtained in Eq. \eqref{apeq:starting_matrix} and conduct the QR decomposition
\begin{equation}
    Q_1R_1=T_1Q_0.
\end{equation}
Then this process is repeated recursively so that
\begin{equation}
    Q_nR_n=T_nQ_{n-1},
\end{equation}
where $Q_n$ is a unitary matrix and $R_n$ is upper triangular. Practically, the matrix product in Eq. \eqref{apeq:starting_matrix} to obtain the initial unitary matrix $Q_0$ can be calculated similarly by performing recursive QR decomposition. As a result,  $\mathcal{R}_n$  in Eq. \eqref{apeq:qr1} is equivalent to $    \mathcal{R}_n=\prod_{i=1}^nR_{i}$ and we obtain
\begin{equation}
    \mathcal{S}_{k,n}=\prod_{i=1}^n|(R_{i})_{k,k}|.
\end{equation}

Due to the central limit theorem, $\log\mathcal{S}_{k,n}=\sum_{i=1}^n\log|(R_{i})_{k,k}|$ behaves like a Gaussian random variable centered at $n\lambda_k$, as shown in Figs.\ref{sfig:distribution}(a-c) for $k=2$. Finally, the LEs can be extracted from these distributions. 
\subsection{The off-diagonal elements of non-Hermitian disordered Green's function}
\begin{figure}
    \centering
    \includegraphics{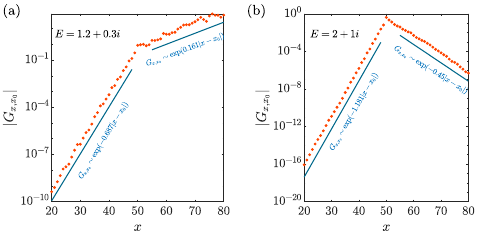}
    \caption{$|G_{x,x_0}(E)|$ for different energies $E$.  The model is $H_2$ in the main text, with the parameters being $t_{-2}=t_2=0.1$, $t_{-1}=1.2$,  $ t_1=0.3$, and $t_0=0.0$. The random potential $V$ is taken from the uniform distribution in the interval $[-0.6,0.6]$. In numerical calculations, we set $L=100$ and $x_0=50$. The numerical results for a specific disorder realization are represented by orange dots, while the blue lines show the theoretical scaling behaviors predicted by Eq. \eqref{apeq:GF_Scaling}. The LEs in these plots are calculated by Eq.\eqref{apeq:lypunov}.}
    \label{sfig:green}
\end{figure}
In the non-Hermitian system without disorders, the non-Hermitian Green's function under open boundary conditions is expressed by \cite{xue2021simple}
\begin{equation}
    \braket{x|\frac{1}{E-H_0}|x_0}=\int_{\text{GBZ}}\frac{\mathrm{d}\beta}{2\pi i\beta}\beta^{x-x_0}\frac{1}{E-h_0(\beta)}.
\end{equation}
$H_0\equiv H_2|_{V_i=0}$ is the translationally invariant part of $H_2$ in the main text, which is generated by the non-Bloch Hamiltonian $\sum_{n=-2}^{2}t_n\beta^n$ with $t_n$ being the hopping elements. The contour integral is performed on the generalized Brillouin zone (GBZ) \cite{xue2021simple}. As a result, the asymptotic behaviors of the OBC Green's function are given by
\begin{equation}\label{apeq:GF_clean}
     \braket{x|\frac{1}{E-H_0}|x_0}\sim\begin{cases}
         |\beta_2(E)|^{x-x_0}, \quad &x\gg x_0;\\
          |\beta_3(E)|^{-|x-x_0|}, \quad &x\ll x_0.        
     \end{cases}
\end{equation}
Here, $|\beta_1(E)|\le|\beta_2(E)|\le|\beta_3(E)|\le|\beta_4(E)|$ are the roots of the characteristic equation $E=h_0(\beta)$.  It is worth noting that the middle two roots $\beta_2(E)$ and $\beta_3(E)$ describe the scaling behaviors of OBC Green's function in this non-Hermitian system. 

In the disordered non-Hermitian system described by $H_2$, the role of $\beta_k(E)$  is replaced by these LEs $\lambda_k$. Therefore, we expect that the off-diagonal scaling properties of non-Hermitian disordered Green's function under open boundary conditions should be revealed by the LEs $\lambda_2$ and $\lambda_3$ through the following expression:
\begin{equation}\label{apeq:GF_Scaling}
     \braket{x|\frac{1}{E-H_2}|x_0}\sim\begin{cases}
        e^{\lambda_2|x-x_0|}, \quad &x\gg x_0;\\
          e^{-\lambda_3|x-x_0|}, \quad &x\ll x_0.        
     \end{cases}
\end{equation}
This result is evident by numerical investigations. The examples are shown in Fig.  \ref{sfig:green}. This formula resembles Eq. \eqref{apeq:GF_clean} where the random potential is turned off.

In a generic disordered non-Hermitian system $H$ with the hopping range from both directions being $M$, the result in Eq. \eqref{apeq:GF_Scaling} is further extended to
\begin{equation}
     \braket{x|\frac{1}{E-H }|x_0}\sim\begin{cases}
        e^{\lambda_M|x-x_0|}, \quad &x\gg x_0;\\
          e^{-\lambda_{M+1}|x-x_0|}, \quad & x\ll x_0.        
     \end{cases}
\end{equation}
In the above, $\lambda_{M}$ and $\lambda_{M+1}$ are the middle two LEs of the random transfer matrix to the right direction.\\

 \subsection{The calculation of cumulant generating functions }

Based on the finite-$n$ distribution of $\mathcal{S}_{k,n}$ [Fig.\ref{sfig:distribution}], we can define the cumulant generating function as
\begin{equation}
    \Lambda_{k,n}(r)=\frac{1}{n}\log\mathbb{E}[{(\mathcal{S}_{k,n})^r}],
\end{equation}
where $\mathbb{E}[\cdot]$ is the average over disorder realizations. Then the cumulant generating function for infinite $n$ is defined as 
\begin{equation}
    \Lambda_k(r)=\lim\limits_{n\to\infty} \Lambda_{k,n}(r).
\end{equation}

It is worth noting that the subleading correction for the finite-$n$ cumulant generating function is given by \begin{equation}
    \Lambda_{k,n}(r)=\Lambda_{k}(r)+\frac{a}{n}+o(\frac{1}{n}).
\end{equation}
Therefore, we can extract $\Lambda_k(r)$ by linear fitting of finite-$n$ data, which is shown in Fig. \ref{sfig:distribution}(d). The resulting $\Lambda_k(r)$ with $k=2$ encodes the scaling behaviors of $G_m(E)$ presented in Fig.  2 of the main text. 

To be concrete, we can extract the phase boundary between the algebraic region and the exponential region by locating the energies $E$ satisfying $\Lambda_2(1)=0$. The exponents of $G_m(E)$ in the exponential region are also obtained by calculating $\Lambda_2(1)$. In addition, the scaling factors of $G_m(E)$ in the algebraic region are found by solving $\Lambda_2(r)=0$. Moreover, the phase boundary between the algebraic region and the bounded region is provided by the critical energies without a nonzero root of $\Lambda_2(r)=0$. This condition is also equivalent to $\max(\log\mathcal{S}_{2,n})=0$, which turns out to be the envelope of all possible Bloch spectrum $h_0(e^{ik})+V$ with $k\in[0,2\pi)$ and $V$ taken from the distribution of the random potential.   Here, $h_0(e^{ik})$ is the Bloch Hamiltonian of the translationally invariant part of $H_2$ in the main text.

\bibliography{GF}